\documentclass{juliacon}
\usepackage[autonum, blackhypersetup]{shortex}
\usepackage{xcolor}
\usepackage[font=scriptsize]{subcaption}
\usepackage{forest}

\definecolor{julia1}{rgb}{0, 0.60, 0.98}
\definecolor{julia2}{rgb}{0.89, 0.43, 0.28}
\definecolor{julia3}{rgb}{0.24, 0.64, 0.30}
\definecolor{julia4}{rgb}{0.76, 0.44, 0.82}
\definecolor{julia5}{rgb}{0.67, 0.55, 0.09}
\definecolor{julia6}{HTML}{00aaae}
\definecolor{julia7}{HTML}{ed5e93}
\definecolor{julia8}{HTML}{c68225}

\begin{document}


\title{Pigeons.jl: Distributed sampling from intractable distributions}

\author[1]{Nikola Surjanovic}
\author[1]{Miguel Biron-Lattes}
\author[2]{Paul Tiede}
\author[3]{Saifuddin Syed}
\author[1]{Trevor Campbell}
\author[1]{Alexandre Bouchard-Côté}
\affil[1]{University of British Columbia}
\affil[2]{Harvard University}
\affil[3]{University of Oxford}


\hypersetup{
pdftitle = {My JuliaCon proceeding},
pdfsubject = {JuliaCon 2019 Proceedings},
pdfauthor = {1st author, 2nd author, 3rd author},
pdfkeywords = {Julia, Optimization, Game theory, Compiler},
}

\maketitle

\begin{abstract}

We introduce a software package, Pigeons.jl, that provides a way to leverage distributed computation to 
obtain samples from complicated probability distributions, such as 
multimodal posteriors arising in Bayesian inference and high-dimensional 
distributions in statistical mechanics. Pigeons.jl provides simple APIs to 
perform such computations single-threaded, multi-threaded,
and/or distributed over thousands of MPI-communicating machines.
In addition, Pigeons.jl guarantees a property that we call strong parallelism invariance: the 
output for a given seed is identical irrespective of the number of threads and processes, 
which is crucial for scientific reproducibility and software validation. 
We describe the key features of Pigeons.jl and the approach taken to implement 
a distributed and randomized algorithm that satisfies strong parallelism invariance.

\end{abstract}

\section{Introduction}
In many scientific application domains, the ability to obtain samples from a 
distribution is crucial. 
For instance, sampling methods have been used to discover magnetic polarization 
in the black hole of galaxy M87 \cite{akiyama2021seven}
and to image the Sagittarius A* black hole \cite{akiyama2022first}.
They have also been used to 
model the evolution of single-cell cancer genomes \cite{salehi2021clonal}, 
infer plasma dynamics inside nuclear fusion reactors \cite{gota2021overview}, 
and to identify gerrymandering in Georgia's 2021 congressional districting plan 
\cite{zhao2022mathematically}.
Similarly, evaluating high-dimensional integrals or sums over complicated 
combinatorial spaces are related tasks that can also be solved with sampling 
via Markov chain Monte Carlo (MCMC) methods. 
However, such calculations can often be bottlenecks in the scientific process, with 
simulations that can last days or even weeks to finish. 

\medskip
Pigeons.jl\footnote{The source code for Pigeons.jl can be found at 
\url{https://github.com/Julia-Tempering/Pigeons.jl}. 
Pigeons.jl v0.2.0, used in the example scripts below, is currently compatible with Julia 1.8+.}
enables users to sample efficiently from high-dimensional and complex distributions 
and solve integration problems by 
implementing state-of-the-art sampling algorithms \cite{syed2021nrpt,surjanovic2022vpt} 
that leverage distributed computation. Its simple API allows users to perform such computations 
single-threaded, multi-threaded, and/or distributed over thousands of MPI-communicating 
machines. Further, it comes with guarantees on strong parallelism invariance wherein 
the output for a given seed is \emph{identical} irrespective 
of the number of threads or processes. 
Such a level of reproducibility is rare in distributed software but of 
great use for the purposes of debugging in the context of sampling algorithms, 
which produce stochastic output.
Specifically, Pigeons.jl is designed to be suitable and yield reproducible output for:
\begin{enumerate}
    \item one machine running on one thread;
    \item one machine running on several threads;
    \item several machines running, each using one thread, and
    \item several machines running, each using several threads.
\end{enumerate}

\subsection{Problem formulation}
We describe the class of problems that can be approached using Pigeons.jl.
Let $\pi(x)$ denote a probability density
\footnote{This density may also be a probability mass function 
(discrete variables). We also allow for a combination of discrete/continuous variables.} 
called the \emph{target}. 
In many problems, e.g. in Bayesian statistics, the density $\pi$ is typically 
known only up to a normalizing constant, 
\[
\label{eq:normalizing_constant}
  \pi(x) = \frac{\gamma(x)}{Z}, \qquad Z = \int \gamma(x) \, \dee x,
\]  
where $\gamma$ can be evaluated pointwise but $Z$ is typically unknown.
Pigeons.jl takes as input the function $\gamma$.

\medskip 
The output of Pigeons.jl can be used for two main tasks:
\begin{enumerate}
    \item Approximating integrals of the form $\int f(x) \pi(x) \, \dee x$.  
    For example, the choice $f(x) = x$ computes the mean and 
    $f(x) = I[x \in A]$ computes the probability of $A$ under $\pi$,
    where $I[\cdot]$ denotes the indicator function.

    \item Approximating the value of the normalization constant $Z$. For 
    example, in Bayesian statistics, this corresponds to the 
    marginal likelihood, which can be used for model selection. 
\end{enumerate}
Its implementation shines compared to traditional sampling packages in the 
following scenarios:
\begin{itemize}
    \item When the target density $\pi$ is challenging due to a complex structure 
    (e.g., high-dimensional, multi-modal, etc.).
    
    \item When the user needs not only $\int f(x) \pi(x) \, \dee x$ but also
    the normalization constant $Z$. 
    Many existing tools focus on the former and struggle or fail to do the latter. 
    
    \item When the target distribution $\pi$ is defined over a non-standard space, 
    e.g. a combinatorial object such as a phylogenetic tree.  
\end{itemize}

\subsection{What is of interest to the general Julia developer?}
Ensuring code correctness at the intersection of randomized, parallel, and 
distributed algorithms is a challenge. To address this challenge, we designed Pigeons.jl 
based on a principle that we refer to as \textit{strong parallelism invariance}
(SPI).
Namely, the output of Pigeons.jl is completely invariant to the number of machines 
and/or threads.
Without explicitly keeping SPI in mind during software construction, 
the (random) output of the algorithm is only guaranteed to have the same distribution.
This is a much weaker guarantee that, in particular, makes debugging difficult.
However, with our notion of SPI we make debugging and 
software validation considerably easier. This is because the developer can 
first focus on the fully serial randomized algorithm, and then use it as an 
easy-to-compare gold-standard reference for parallel/distributed implementations. 
This strategy is used extensively in Pigeons.jl to ensure correctness. In contrast, 
testing equality in distribution, while possible (e.g., see \cite{geweke2004getting}), 
incurs additional false negatives due to statistical error.

\medskip 
The general Julia developer will be interested in: 
\begin{itemize}
    \item The main causes of a violation of strong parallelism invariance that we have identified 
    (\cref{sec:PI_causes})---some of which are specific to Julia---and how we address 
    them in Pigeons.jl.

    \item The SplittableRandoms.jl\footnote{\url{https://github.com/Julia-Tempering/SplittableRandoms.jl}}
    package that was developed by our team 
    to achieve strong parallelism invariance in Pigeons.jl (\cref{sec:splittable_randoms}).
\end{itemize}

\section{Examples}
In this section we present a set of minimal examples 
that demonstrate how to use Pigeons.jl for sampling. 
For further reading, we also direct readers to our growing list of examples, which can be found at 
\url{https://github.com/Julia-Tempering/Pigeons.jl/tree/main/examples}.
We begin by installing the latest official release of Pigeons.jl: 
\begin{lstlisting}[language = Julia]
using Pkg; Pkg.add("Pigeons")
\end{lstlisting}

\subsection{Targets}
To use Pigeons.jl, we must specify a target distribution, given by $\gamma$ 
in \cref{eq:normalizing_constant}.
Numerous possible types of target distributions are supported, including 
custom probability densities (specified up to a normalizing constant) written in Julia.
We also allow to interface with models written in common probabilistic programming 
languages, including:
\begin{itemize}
    \item Turing.jl \cite{ge2018turing} models (\texttt{TuringLogPotential})
    \item Stan \cite{carpenter2017stan} models (\texttt{StanLogPotential})
    \item Comrade.jl\footnote{\url{https://github.com/ptiede/Comrade.jl}} 
      models for black hole imaging (\texttt{ComradeLogPotential})
    \item Non-Julian models with foreign-language Markov chain Monte Carlo (MCMC) code 
    (e.g. Blang \cite{bouchard2022blang} code for phylogenetic inference over combinatorial spaces) 
\end{itemize}
Additional targets are currently being accommodated and will be 
introduced to Pigeons.jl in the near future.

\begin{figure}[t]
    \centering 
    \includegraphics[width=0.5\textwidth]{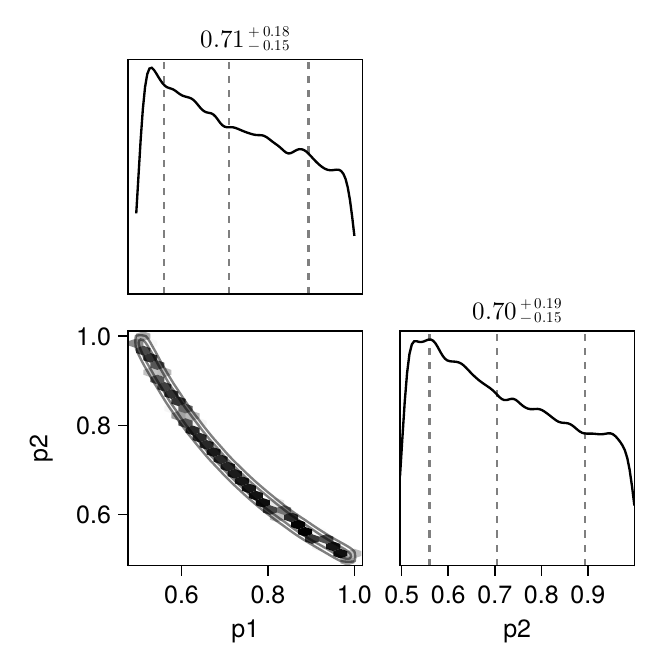}
    \caption{Posterior distribution for the model given by \cref{eq:coinflip} 
    with $n=100,000$ coin flips and $y=50,000$ observed heads, 
    estimated using $2^{17}$ samples from Pigeons.jl. 
    We present the pairwise plot for $p_1$ and $p_2$, as well as the estimated 
    densities of the marginal of the posterior for each of the two parameters.
    Note that because the model is non-identifiable, as we collect more data the 
    posterior distribution concentrates around the curve $p_1 p_2 = 0.5$, instead 
    of a single point, assuming that the true probability of observing heads 
    is 0.5.}
    \label{fig:coinflip_posterior}
\end{figure}

\medskip 
In what follows, we demonstrate how to use Pigeons with a Julia Turing model 
applied to a non-identifiable ``coinflip'' data set, modified from the example
at \url{https://turing.ml/v0.22/docs/using-turing/quick-start}. 
The Bayesian model can be formulated as 
\[
  \label{eq:coinflip}
  p_1, p_2 &\stackrel{iid}{\sim} U(0, 1) \\    
  Y \mid p_1, p_2 &\sim \text{Binomial}(n, p_1 p_2).
\]
The random variable $Y$ is the number of heads observed on $n$ coin flips
where the probability of heads is $p_1 p_2$.  
This model is non-identifiable, meaning that it is not possible to distinguish 
the effects of the two different parameters $p_1$ and $p_2$. As a consequence, 
the target distribution exhibits a complicated structure, as displayed in 
\cref{fig:coinflip_posterior}.
The density of interest corresponding to this model is 
\[
  \pi(p_1, p_2) &= \gamma(p_1, p_2)/Z,
\] 
where 
\[ 
  \gamma(p_1, p_2) &= 
    {n \choose y} (p_1 p_2)^y (1-p_1 p_2)^{n-y} I[p_1, p_2 \in [0,1]] 
    \label{eq:coinflip_density} \\
  Z &= \int_0^1 \int_0^1 \gamma(p_1, p_2) \, \dee p_1 \, \dee p_2. \label{eq:coinflip_normalization}
\]
The distribution $\pi$ is also known as the \emph{posterior distribution} in 
Bayesian statistics.

\medskip 
Suppose that we perform $n=100,000$ coin tosses and observe 
$y=50,000$ heads.
We would like to obtain samples from our posterior, $\pi$, having collected this data.
We begin by installing Turing
\begin{lstlisting}[language = Julia]
Pkg.add("Turing")
\end{lstlisting}
and then defining our Turing model and storing it in the variable \texttt{model}:
\begin{lstlisting}[language = Julia]
using Turing
@model function coinflip(n, y)
    p1 ~ Uniform(0.0, 1.0)
    p2 ~ Uniform(0.0, 1.0)
    y ~ Binomial(n, p1 * p2)
    return y
end
model = coinflip(100000, 50000)
\end{lstlisting}

From here, it is straightforward to sample from the density given by 
\cref{eq:coinflip_density} up to a normalizing constant.
We use non-reversible parallel tempering \cite{syed2021nrpt} (PT), Pigeons.jl's 
state-of-the-art sampling algorithm, to sample from the target distribution.
PT comes with several tuning parameters and \cite{syed2021nrpt} describe how 
to select these parameters effectively, which Pigeons.jl implements under the hood.
We also specify that we would like to store 
the obtained samples in memory to be able to produce trace-plots, as well 
as some basic online summary statistics of the target distribution and useful 
diagnostic output by specifying 
\texttt{record = [traces, online, round\_trip, Pigeons.timing\_extrema, Pigeons.allocation\_extrema]}.
It is also possible to leave the \texttt{record} argument empty and reasonable defaults 
will be selected for the user.
The code below runs Pigeons.jl on one machine with one thread.
We use the default values for most settings, however 
we explain later how one can obtain improved performance by setting 
arguments more carefully (see \cref{sec:additional_options}).
\begin{lstlisting}[language = Julia]
using Pigeons
pt = pigeons(
    target = TuringLogPotential(model), 
    record = [
        traces, online, round_trip, 
        Pigeons.timing_extrema, 
        Pigeons.allocation_extrema])
\end{lstlisting}
Note that to convert the Turing model into an appropriate Pigeons.jl target for sampling, 
we pass the model as an argument to \texttt{TuringLogPotential()}.
Once we have stored the PT output in the variable \texttt{pt} we can 
access the results, as described in the following section. 
The standard output after running the above code chunk is displayed in \cref{fig:standard_out}
and explained in the next section. 
For purposes of comparison, we also run a traditional 
(single-chain Markov chain Monte Carlo) method. 

\begin{figure*}[t]
    \centering
    \begin{lstlisting}
          --------------------------------------------------------------------------------------
            #scans    restarts      Λ        time(s)    allc(B)  log(Z₁/Z₀)   min(α)     mean(α)
          ---------- ---------- ---------- ---------- ---------- ---------- ---------- --------- 
                  2          0       1.04      0.383   3.48e+07  -4.24e+03          0      0.885
                  4          0       4.06    0.00287   1.79e+06      -16.3   4.63e-06      0.549
                  8          0       3.49    0.00622   3.55e+06      -12.1      0.215      0.612
                 16          0       2.68     0.0161   7.46e+06      -10.2      0.518      0.703
                 32          0       4.29     0.0353   1.37e+07      -11.8      0.222      0.524
                 64          3       3.17     0.0699   2.86e+07      -11.5      0.529      0.648
                128          8       3.56      0.139   5.53e+07      -11.5      0.523      0.605
                256         12       3.38      0.241    1.1e+08      -11.6      0.526      0.625
                512         37       3.48      0.473   2.22e+08        -12      0.527      0.614
           1.02e+03         77       3.55      0.895   4.46e+08      -11.8      0.571      0.605
          --------------------------------------------------------------------------------------
    \end{lstlisting}
    \caption{Standard output provided by Pigeons.jl. 
    Rows indicate tuning rounds of the PT algorithm with an exponentially increasing number 
    of PT iterations (\texttt{\#scans}). Columns indicate various useful diagnostics, 
    such as the number of allocations per round, time (in seconds), and estimates 
    of the log of the normalization constants. 
    The output is described in greater detail in \cref{sec:PT_diagnostics}
    and has been modified to exclude columns that are not described in the paper.}
    \label{fig:standard_out}
\end{figure*}

\subsubsection{Other targets}
As mentioned previously, it is also possible to specify targets with custom 
probability densities, as well as Stan and Turing models. 
Additionally, suppose we have some code implementing vanilla MCMC, written 
in an arbitrary ``foreign'' language such as C++, Python, R, Java, etc. 
Surprisingly, it is very simple to bridge such code with Pigeons.jl. 
The only requirement on the foreign language is that it supports 
reading the standard input and writing to the standard output, 
as described in our online documentation.

\subsection{Outputs}
Pigeons.jl provides many useful types of output, such as: plots of samples from 
the distribution, estimates of normalization constants, summary statistics of 
the target distribution, and various other diagnostics. 
We describe several examples of possible output below.

\subsubsection{Standard output}
An example of the standard output provided by Pigeons.jl is displayed in \cref{fig:standard_out}. 
Each row of the table in the output indicates a new tuning round in parallel tempering, 
with the \texttt{\#scans} column indicating the number of scans/samples in 
that tuning round. 
During these tuning rounds, Pigeons.jl searches for optimal values of certain PT 
tuning parameters. Other outputs include: 
\begin{itemize}
    \item \texttt{restarts}: a higher number is better. 
    Informally, PT performs well at sampling from high-dimensional and/or multi-modal 
    distributions by pushing samples from an easy-to-sample distribution (the reference) 
    to the more difficult target distribution. 
    A tempered restart happens when a sample from the reference successfully 
    percolates to the target. (See the subsequent sections for a more detailed description 
    of parallel tempering.) 
    When the reference supports i.i.d.~sampling, tempered restarts 
    can enable large jumps in the state space. 

    \item \texttt{$\Lambda$}: the global communication barrier, as described in \cite{syed2021nrpt}, 
    which measures the inherent difficulty of the sampling problem.
    A rule of thumb to configure the number of PT chains is also 
    given by \cite{syed2021nrpt}, where they suggest that stable performance 
    should be achieved when the number of chains is set to roughly 2$\Lambda$.
    See \cref{sec:PT_diagnostics} for more information.

    \item \texttt{time} and \texttt{allc}: the time (in seconds) and number of 
    allocations (in bytes) used in each round.

    \item \texttt{$\log(Z_1/Z_0)$}: an estimate of the 
    logarithm of the ratio of normalization constants between the target and the reference. 
    In many cases, $Z_0 = 1$.

    \item \texttt{min($\alpha$)} and \texttt{mean($\alpha$)}: minimum and average swap 
    acceptance rates during the communication phase across the PT chains.
    See \cref{sec:PT} for a description of PT communication.
\end{itemize}

\subsubsection{Plots}
It is straightforward to obtain plots of samples from the target distribution, 
such as trace-plots, pairwise plots, and density plots of the marginals. 

\medskip 
To obtain posterior densities and trace-plots, we first 
make sure that we have the third-party 
MCMCChains.jl\footnote{\url{https://github.com/TuringLang/MCMCChains.jl}}, 
StatsPlots.jl\footnote{\url{https://github.com/JuliaPlots/StatsPlots.jl}}, and 
PlotlyJS.jl\footnote{\url{https://github.com/JuliaPlots/PlotlyJS.jl}} 
packages installed via
\begin{lstlisting}[language=Julia]
Pkg.add("MCMCChains", "StatsPlots", "PlotlyJS")
\end{lstlisting}

\begin{figure*}[t]
    \centering
    \begin{subfigure}{0.45\textwidth}
      \centering
      \includegraphics[width=\textwidth]{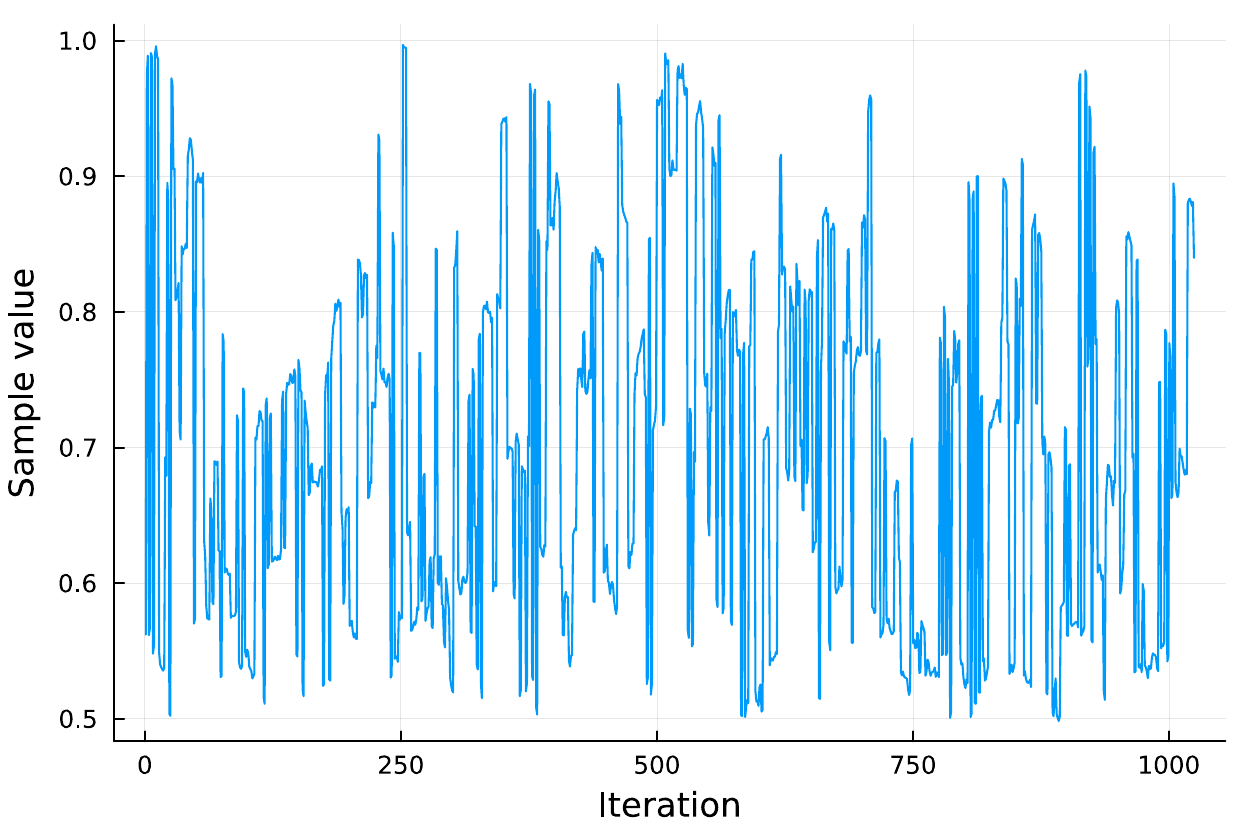}
    \end{subfigure}
    \begin{subfigure}{0.45\textwidth}
      \centering
      \includegraphics[width=\textwidth]{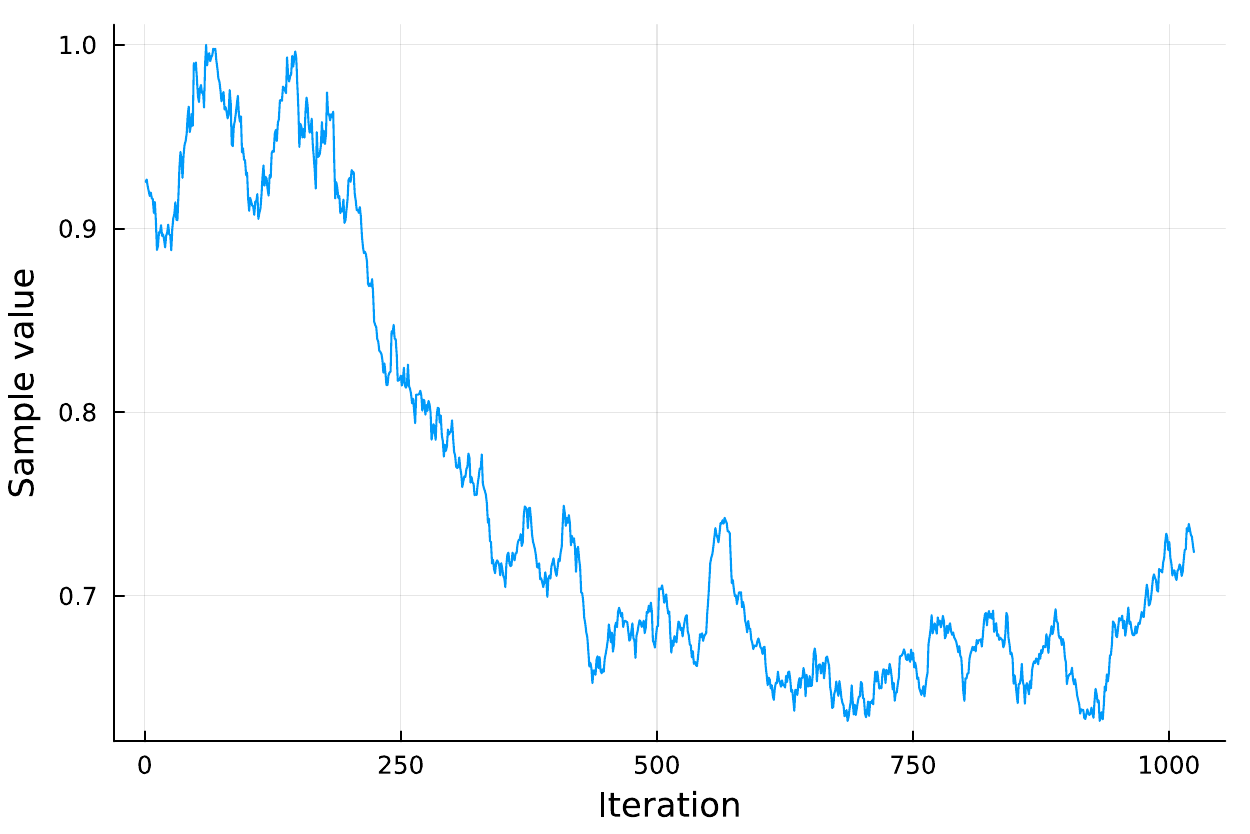}
    \end{subfigure}
    \caption{
        Trace-plots for the first parameter, $p_1$, in the 
        non-identifiable coinflip Turing model.
        \textbf{Left:} Samples from Pigeons.jl using PT with 10 chains. Note that the 
        trace-plot indicates fast mixing/exploration across the state space.
        \textbf{Right:} Single-chain Markov chain Monte Carlo. Note that the 
        trace-plot explores the state space much more slowly when we do not use PT.
    }
    \label{fig:trace_density_plots}
\end{figure*}

With the \texttt{pt} output object from before for our non-identifiable coinflip model, 
we can run the following:
\begin{lstlisting}[language=Julia]
using MCMCChains, StatsPlots, PlotlyJS
plotlyjs()
samples = Chains(
    sample_array(pt), variable_names(pt))
my_plot = StatsPlots.plot(samples)
display(my_plot)
\end{lstlisting}
The output of the above code chunk is an interactive plot that can be zoomed in or out
and exported as an HTML webpage. 
A modified static version of the output for the first parameter, $p_1$, 
is displayed in the top panel of \cref{fig:trace_density_plots},
along with a comparison to the output from a single-chain algorithm in the bottom 
panel of the same figure.

\medskip 
To obtain pair plots, we add the 
PairPlots.jl\footnote{\url{https://github.com/sefffal/PairPlots.jl}}
and CairoMakie.jl\footnote{\url{https://github.com/JuliaPlots/CairoMakie.jl}}
packages:
\begin{lstlisting}[language=Julia]
Pkg.add("PairPlots", "CairoMakie")
\end{lstlisting}
and then run
\begin{lstlisting}[language=Julia]
using PairPlots, CairoMakie
my_plot = PairPlots.pairplot(samples) 
display(my_plot)
\end{lstlisting}
The output of the above code chunk with an increased number of samples 
is displayed in \cref{fig:coinflip_posterior}.
\footnote{The code chunk above only works with Julia 1.9.}

\subsubsection{Estimate of normalization constant}
The (typically unknown) constant $Z$ in \cref{eq:normalizing_constant} is 
referred to as the \emph{normalization constant}. 
In many applications, it is useful to approximate this constant. 
For example, in Bayesian statistics, this corresponds to the 
marginal likelihood and can be used for model selection. 

\medskip 
As a side-product of PT, we automatically obtain an approximation to the natural 
logarithm of the normalization constant. This is done automatically using the 
stepping stone estimator \cite{xie2011improving}.
The estimate can be accessed using
\begin{lstlisting}[language=Julia]
stepping_stone(pt)
\end{lstlisting}
In the case of the normalization constant given by \cref{eq:coinflip_normalization} 
for $n=100,000$ and $y=50,000$, we can exactly obtain its value as $\log(Z) \approx -11.8794$ 
using a computer algebra system. Note that this is very close to the output provided in 
\cref{fig:standard_out}.

\subsubsection{Online statistics}
\label{sec:online_stats}
Having specified the use of the \texttt{online} recorder in our call to \texttt{pigeons()}, 
we can output some basic summary statistics of the marginals of our target distribution. 
For instance, it is straightforward to estimate the mean and variance of each 
of the marginals of the target with 
\begin{lstlisting}[language=Julia]
using Statistics
mean(pt); var(pt)
\end{lstlisting}
Other constant-memory statistic accumulators are made available in the 
OnlineStats.jl \cite{day2020onlinestats} package. 
To add additional constant-memory statistic accumulators, we 
can register them via \texttt{Pigeons.register\_online\_type()}, as described in our 
online documentation. For instance, we can also compute constant-memory estimates 
of extrema of our distribution.

\subsubsection{Off-memory processing}
When either the dimensionality of the model or the number of samples is large,
the obtained samples may not fit in memory. 
In some cases it may be necessary to store samples to disk if our statistics of 
interest cannot be calculated online and with constant-memory
(see \cref{sec:online_stats}).
We show here how to save samples to disk when Pigeons.jl is run on a single 
machine. A similar interface can be used over MPI. 

\medskip 
First, we make sure that we set \texttt{checkpoint = true}, which saves a 
snapshot at the end of each round in the directory \texttt{results/all/<unique directory>}
and is symlinked to \texttt{results/latest}.
Second, we make sure that we use the \texttt{disk} recorder 
by setting \texttt{record = [disk]}, along with possibly any other desired recorders.
Accessing the samples from disk can then be achieved in a simple way using the Pigeons.jl 
function \texttt{process\_sample()}.

\subsubsection{PT diagnostics}
\label{sec:PT_diagnostics}
We describe how to produce some key parallel tempering diagnostics from 
\cite{syed2021nrpt}.

\medskip 
The global communication barrier, denoted $\Lambda$ in Pigeons.jl output, can be 
used to approximately inform the appropriate number of chains. 
Based on \cite{syed2021nrpt}, stable PT performance should be achieved when  
the number of chains is set to roughly $2\Lambda$. This can be achieved by 
modifying the \texttt{n\_chains} argument in the call to \texttt{pigeons()}. 
The global communication barrier is shown at each round and can also be accessed 
with 
\begin{lstlisting}[language=Julia]
Pigeons.global_barrier(pt)
\end{lstlisting}

The number of restarts per round can be accessed with 
\begin{lstlisting}[language=Julia]
n_tempered_restarts(pt)
\end{lstlisting}
These quantities are also displayed in \cref{fig:standard_out}. 
Many other useful PT diagnostic statistics and plots can be obtained, as described 
in our full documentation.

\subsection{Parallel and distributed PT}
One of the main benefits of Pigeons.jl is that it allows users to easily parallelize 
and/or distribute their PT sampling efforts. We explain how to run MPI locally on 
one machine and also how to use MPI when a cluster is available.

\subsubsection{Running MPI locally}
To run MPI locally on one machine using four MPI processes and one thread per process,
use
\begin{lstlisting}[language = Julia]
pigeons(
    target = TuringLogPotential(model), 
    on = ChildProcess(
            n_local_mpi_processes = 4,
            n_threads = 1))
\end{lstlisting}

\subsubsection{Running MPI on a cluster}
Often, MPI is available via a cluster scheduling system. To run MPI over 
several machines:
\begin{enumerate}
    \item In the cluster login node, follow the Pigeons.jl installation instructions
    in our online documentation. 
    \item Start Julia in the login node, and perform a one-time setup by 
    calling \texttt{Pigeons.setup\_mpi()}.
    \item In the Julia REPL running in the login node, run:
\end{enumerate}
\begin{lstlisting}[language = Julia]
pigeons(
    target = TuringLogPotential(model), 
    n_chains = 1_000,
    on = MPI(n_mpi_processes = 1_000, 
             n_threads = 1))
\end{lstlisting}
The code above will start a distributed PT algorithm with 1,000 chains on 1,000 
MPI processes each using one thread.
Note that for the above code chunks involving \texttt{ChildProcess()} and 
\texttt{MPI()} to work, it may be necessary to specify dependencies in their 
function calls.

\subsection{Additional options}
\label{sec:additional_options}
In the preceding example we only specified the target distribution and let 
Pigeons.jl decide on default values for most other settings of the inference engine. 
There are various settings we can change, including: 
the random seed (\texttt{seed}), the number of PT chains (\texttt{n\_chains}), 
the number of PT tuning rounds/samples (\texttt{n\_rounds}), 
and a variational reference distribution family (\texttt{variational}), among other settings.
For instance, we can run 
\begin{lstlisting}[language = Julia]
pigeons(
    target = TuringLogPotential(model),
    n_rounds = 10, 
    n_chains = 10,
    variational = GaussianReference(),
    seed = 2
)
\end{lstlisting}
which runs PT with the same Turing model target as before and explicitly states 
that we should use 10 PT tuning rounds with 10 chains (described below). 
In the above code chunk we also specify that we would like to use a 
Gaussian variational reference distribution.
That is, the reference distribution is chosen from a multivariate Gaussian family 
that lies as close as possible to the target distribution in order to improve 
the efficiency of PT. We refer readers to \cite{surjanovic2022vpt} for more details.
When only continuous parameters are of interest, we encourage users to consider 
using \texttt{variational = GaussianReference()} and setting \texttt{n\_chains\_variational = 10}, 
for example, as the number of restarts may substantially increase with these settings.

\section{Parallel tempering}
\label{sec:PT}
Pigeons.jl provides an implementation of distributed parallel tempering (PT) described 
in \cite{syed2021nrpt}, which we outline in Algorithm~\ref{alg:distributed_PT}.
This section gives both a brief overview of PT and some details of 
its distributed implementation.

\medskip 
In this section we assume a basic understanding of 
Markov chain Monte Carlo (MCMC) methods.
For readers unfamiliar with MCMC, it is important to know that 
it is a method to obtain approximate samples from a distribution. 
More specifically, it is a class of algorithms for simulating a Markov chain of 
states that look like draws from the target distribution $\pi$.
However, with traditional 
MCMC methods, the samples may be heavily correlated instead of independent 
and may fail to sufficiently explore the 
full space of the distribution (see \cref{fig:bimodal}); 
PT is a method that aims to address these issues.
For a more in-depth review of PT, we refer readers to \cite{surjanovic2022vpt}.

\subsection{Overview of PT}
\begin{figure}[t]
    \centering
    \includegraphics[width=0.3\textwidth]{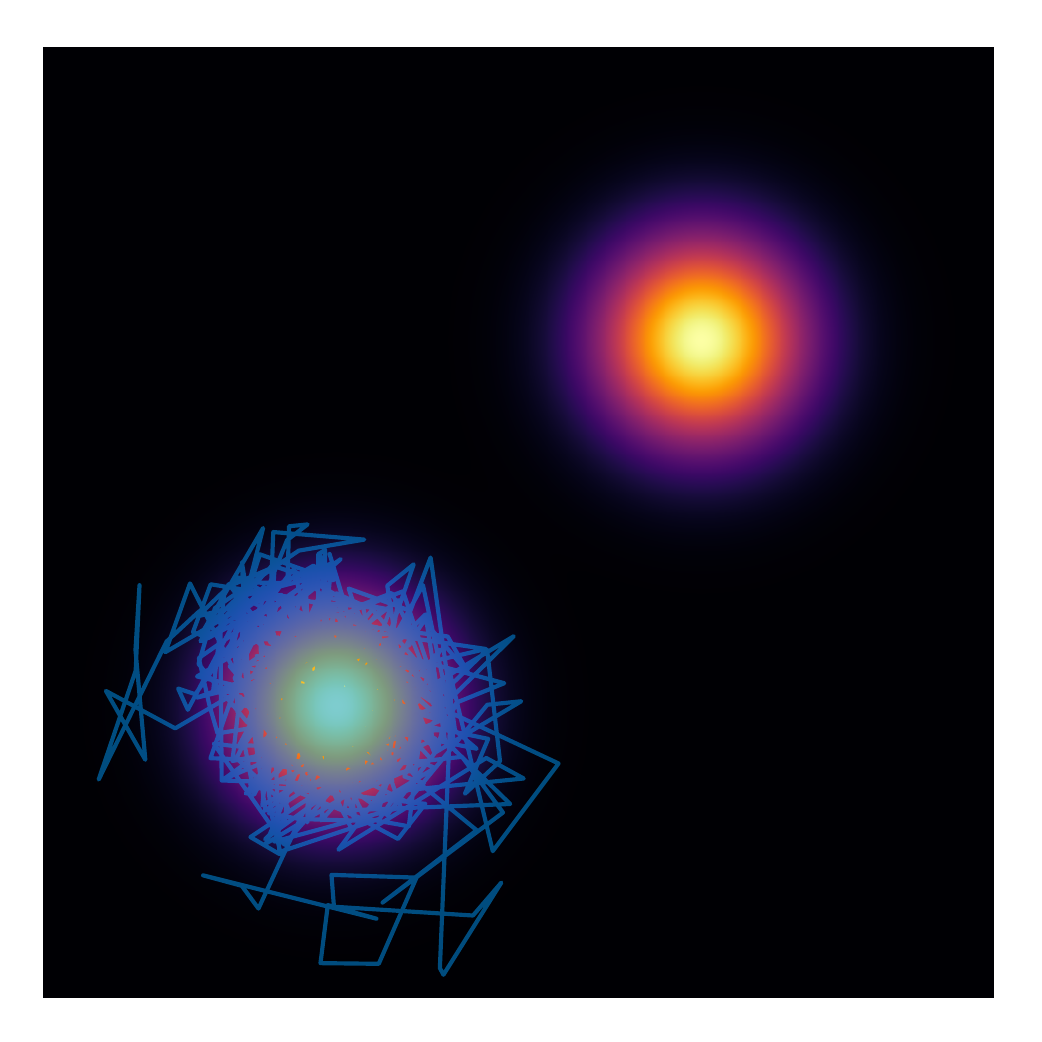}
    \caption{A simple bimodal distribution from which traditional 
    MCMC methods may struggle to obtain samples. Blue lines display output 
    from 1,000 iterations of a Metropolis-Hastings random walk MCMC algorithm.
    The sampler in this figure is visibly trapped in one of the two modes.}
    \label{fig:bimodal}
\end{figure}

Suppose that we would like to estimate integrals involving $\pi$, such as 
$\int f(x) \pi(x) \, \dee x$. 
These integrals may be multivariate and even include combinations of 
continuous and discrete variables (where sums replace integrals in the discrete 
case).
One method is to obtain samples from $\pi$ to approximate such integrals.
Often, the distribution 
$\pi$ can be challenging for traditional MCMC methods---such 
as Metropolis-Hastings, slice sampling, and Hamiltonian Monte Carlo---
because of its structure.
For example, in a bimodal example such as the one illustrated in 
Figure~\ref{fig:bimodal},
traditional methods might remain in one of the two modes for an extremely 
long period of time.

\begin{figure}[t]
    \centering
    \begin{subfigure}{0.15\textwidth}
      \centering
      \includegraphics[width=\textwidth]{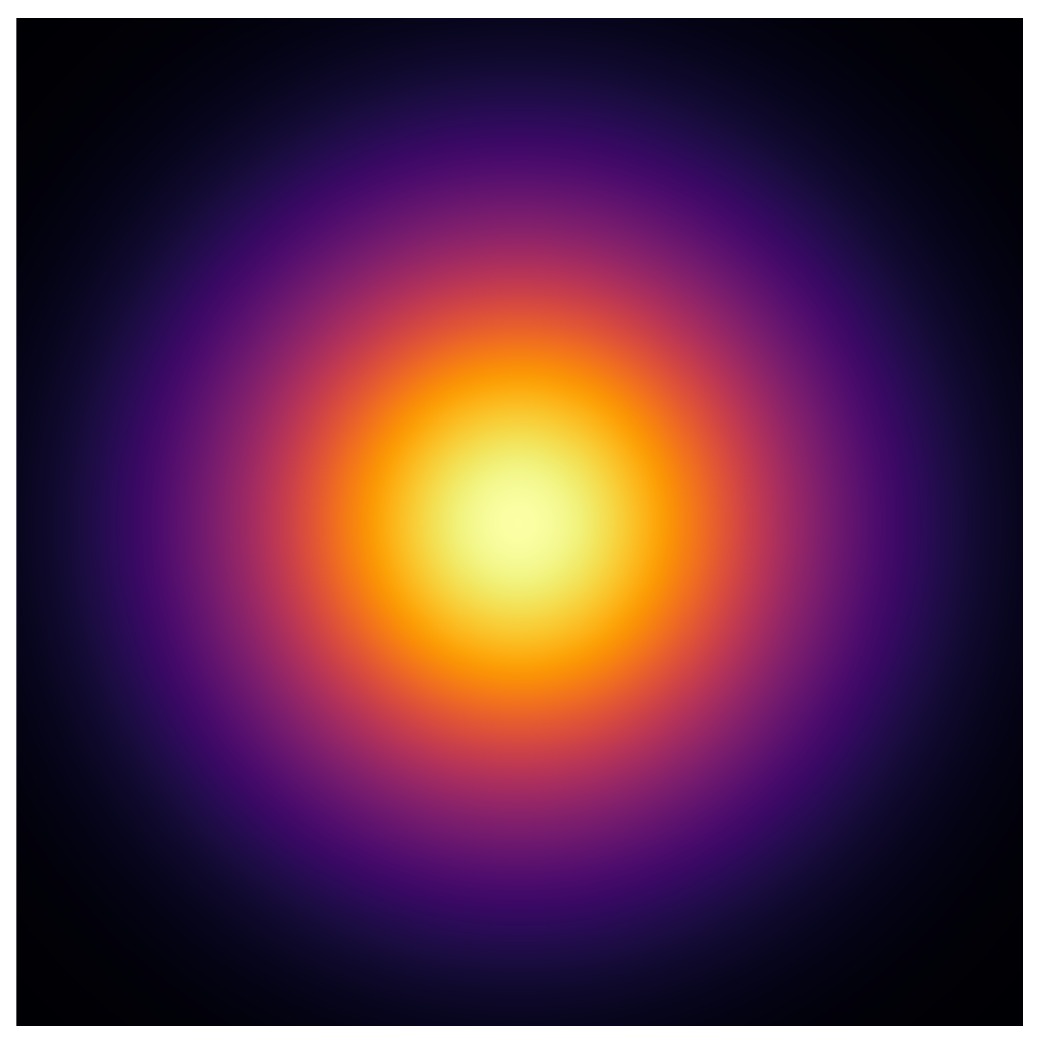}
      \caption*{$\pi_1$}
    \end{subfigure}
    \begin{subfigure}{0.15\textwidth}
      \centering
      \includegraphics[width=\textwidth]{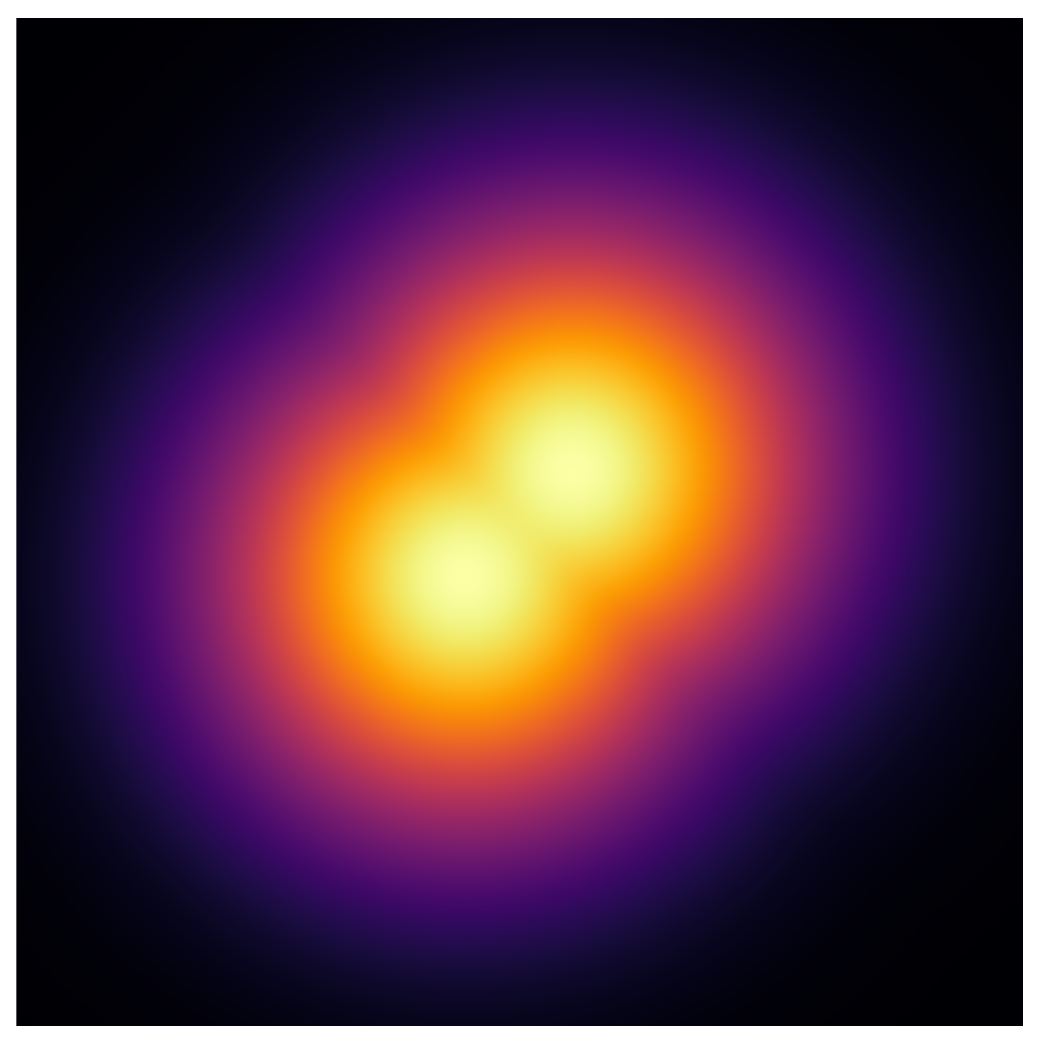}
      \caption*{$\pi_2$}
    \end{subfigure}
    \begin{subfigure}{0.15\textwidth}
      \centering
      \includegraphics[width=\textwidth]{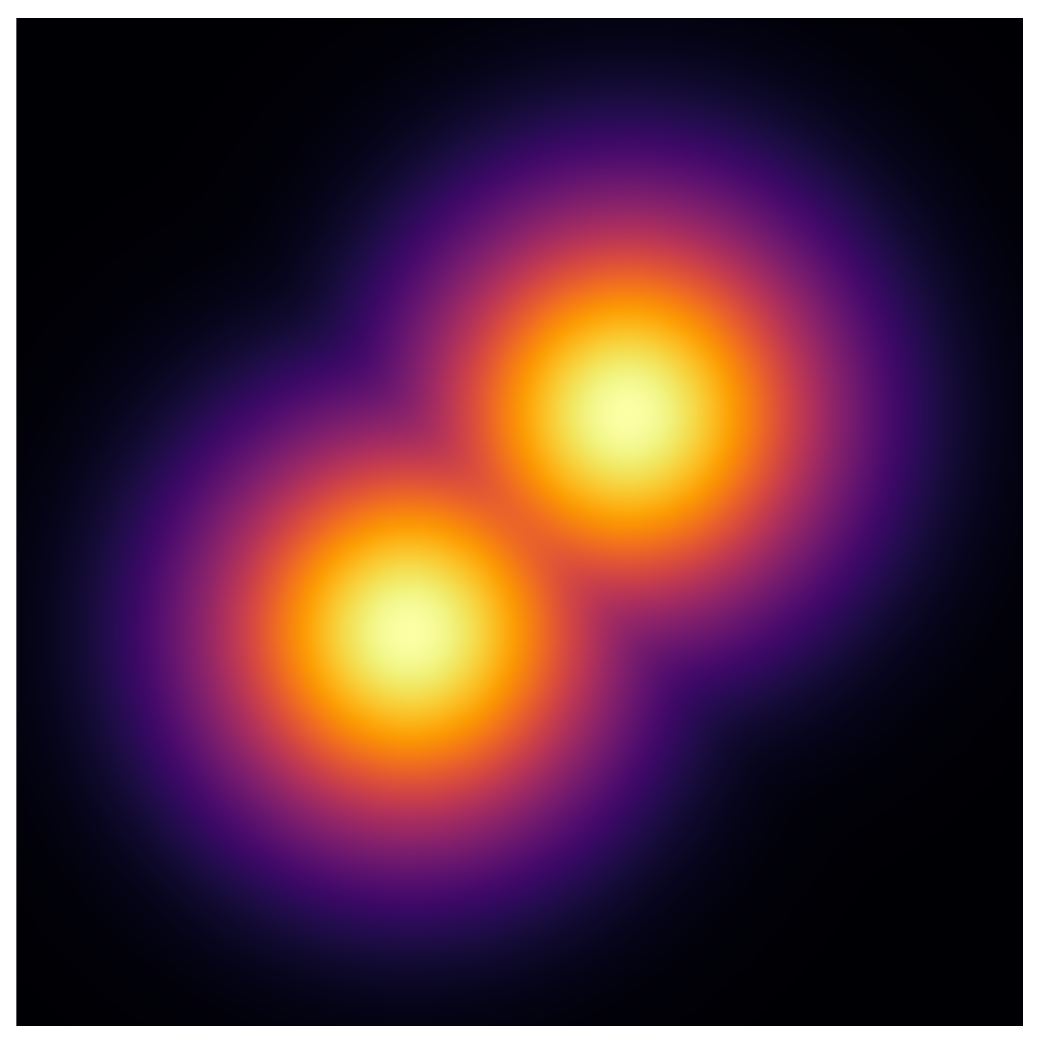}
      \caption*{$\pi_3$}
    \end{subfigure}
    \begin{subfigure}{0.15\textwidth}
      \centering
      \includegraphics[width=\textwidth]{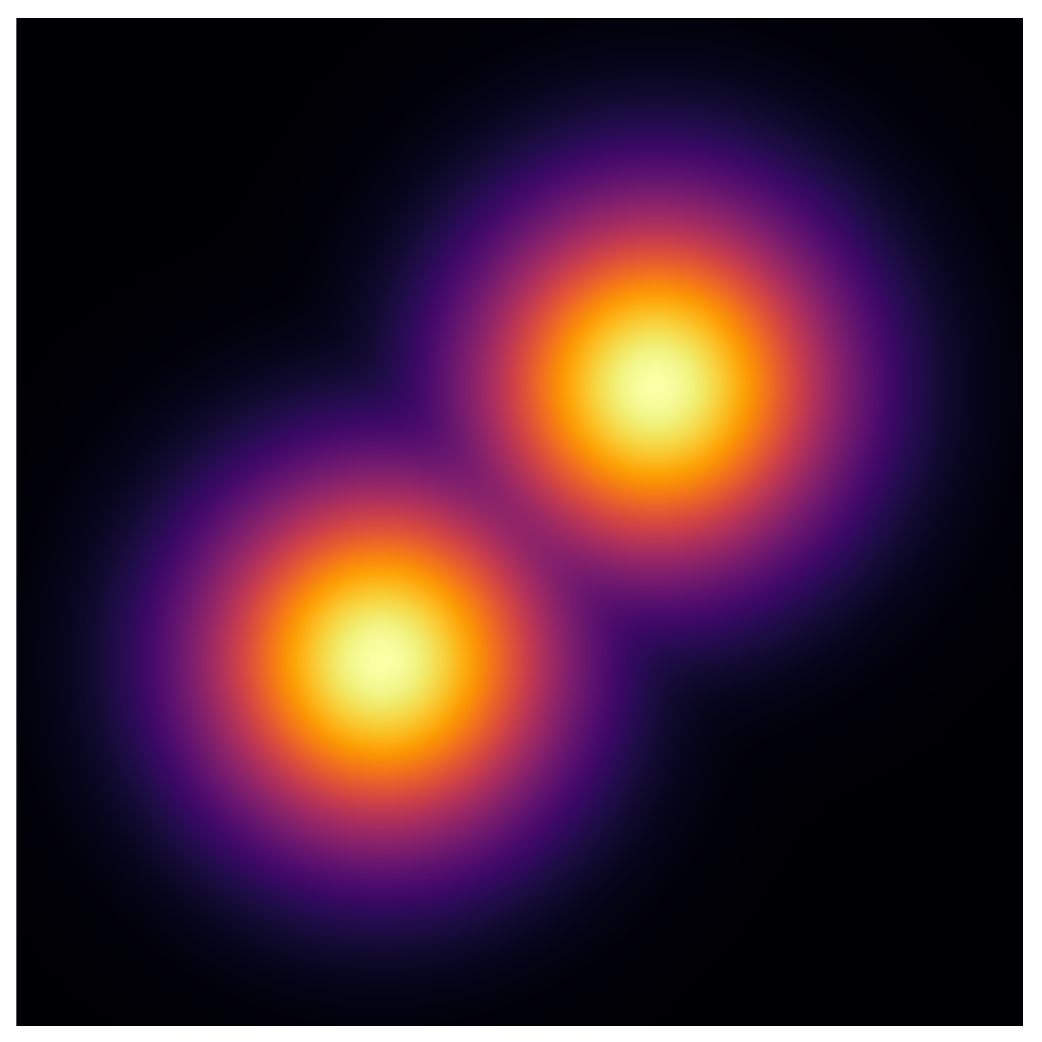}
      \caption*{$\pi_4$}
    \end{subfigure}
    \begin{subfigure}{0.15\textwidth}
      \centering
      \includegraphics[width=\textwidth]{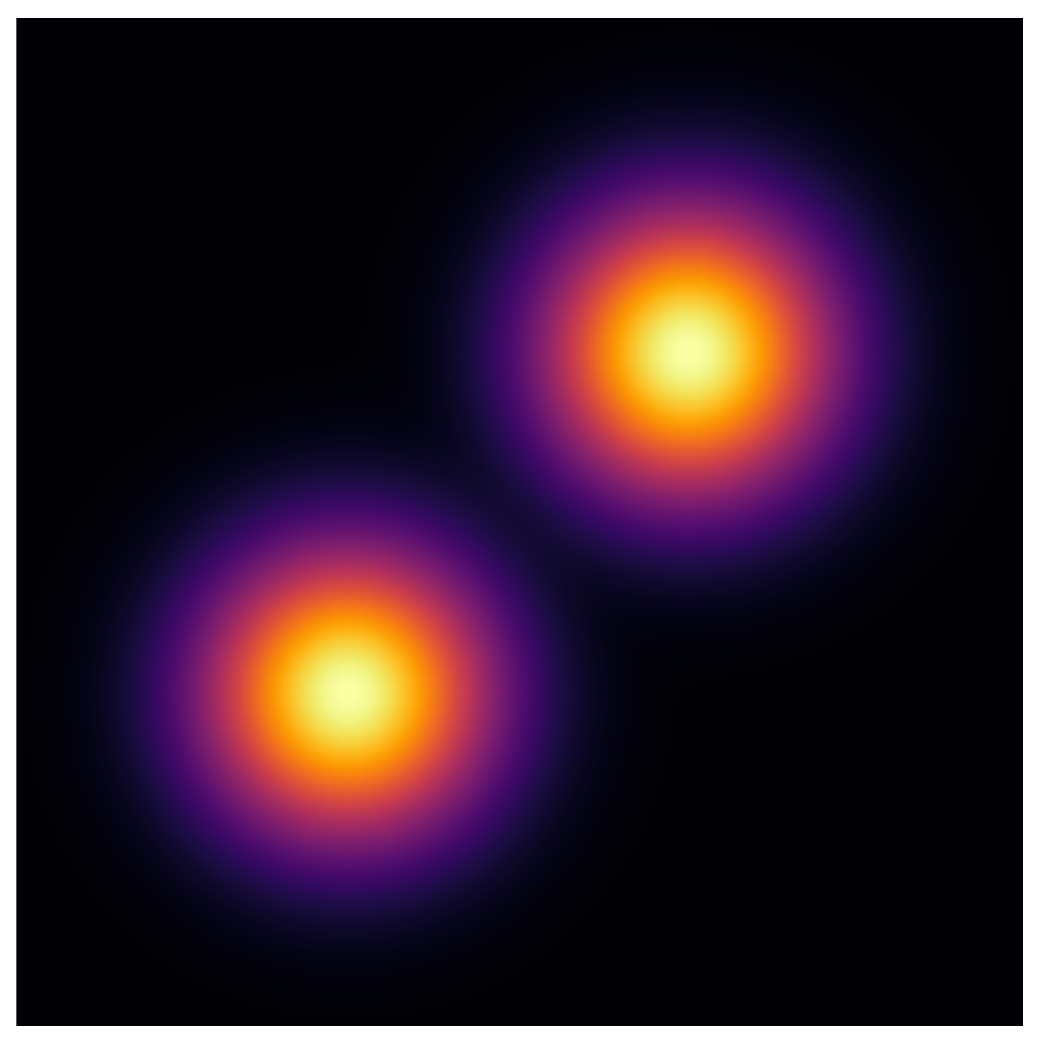}
      \caption*{$\pi_5$}
    \end{subfigure}
    \begin{subfigure}{0.15\textwidth}
      \centering
      \includegraphics[width=\textwidth]{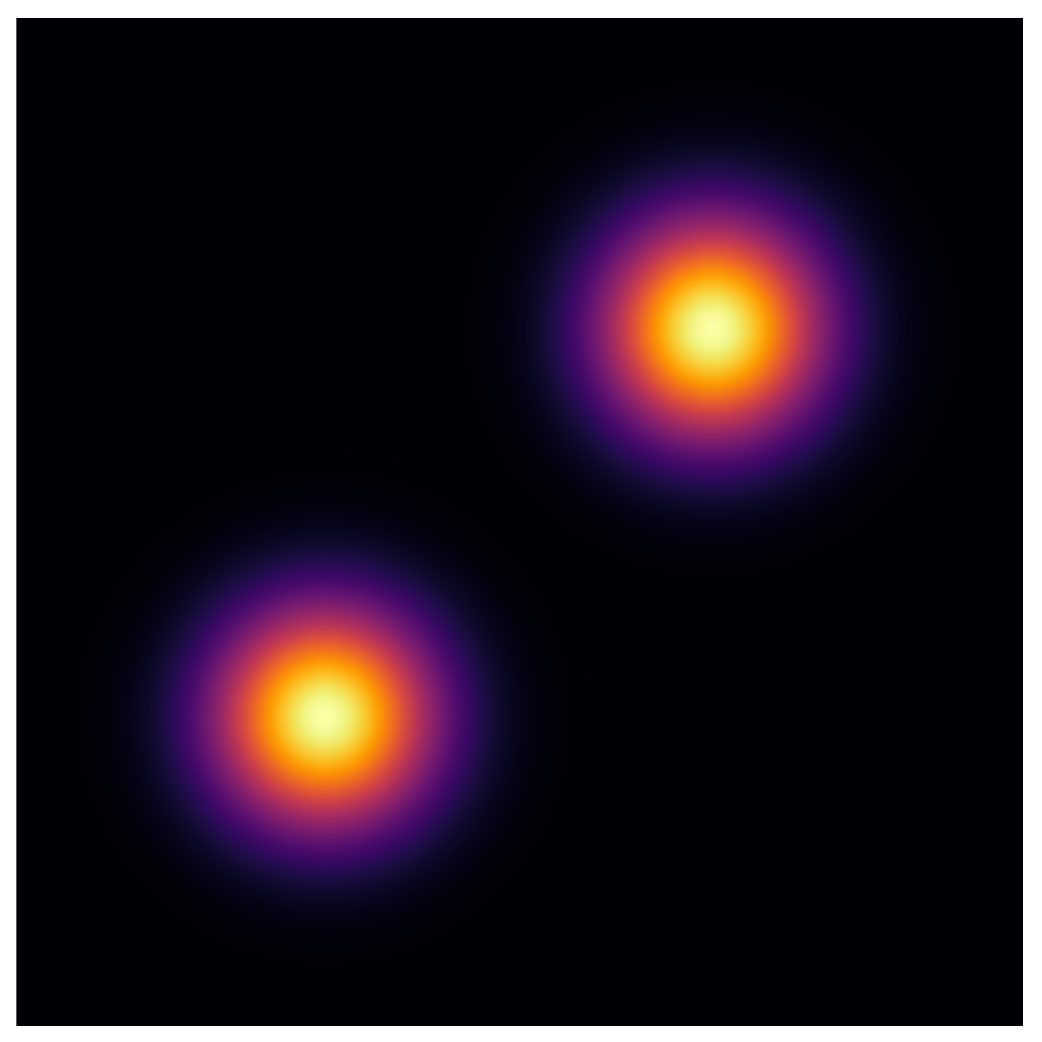}
      \caption*{$\pi_6$}
    \end{subfigure}
    \caption{Heatmaps of six distributions lying on an annealing path 
    from a unimodal reference distribution  
    $\pi_1$, from which it is straightforward to obtain samples, and ending at 
    $\pi_N$, which is in this case the bimodal distribution from 
    Figure~\ref{fig:bimodal}. (Note that in this case the colours between the heatmaps 
    are not directly comparable because the densities of intermediate distributions 
    are not normalized.)}
    \label{fig:path}
\end{figure}

To resolve this issue, PT constructs a sequence of $N$ distributions,  
$\pi_1, \pi_2, \ldots, \pi_N$, where $\pi_N$ is usually equal to $\pi$.
The distributions are chosen so that it is easy to obtain samples from $\pi_1$
with the sampling difficulty increasing as one approches $\pi_N$. An example 
of such a sequence of distributions, referred to as an \textit{annealing path},
is shown in Figure~\ref{fig:path}.

We now turn to explain how this path of distributions can be used to enhance 
sampling from the target distribution, $\pi$. 
PT operates by first obtaining samples from each distribution on the path in parallel 
(referred to as an \textit{exploration phase}). 
Then, samples between adjacent distributions are swapped (referred to as 
a \textit{communication phase}). The communication phase in PT is crucial: it allows 
for the discovery of new regions of the space of the target distribution such as 
the top-right mode of the distribution presented in Figure~\ref{fig:bimodal}.

\subsection{Local exploration and communication}
A struct that is useful for the implementation of PT is the \texttt{Replica}, 
which we will often refer to in the remainder of this section. 
A single \texttt{Replica} struct stores a \texttt{state} variable and a \texttt{chain} 
integer, among other entries. 
At the beginning of PT, $N$ \texttt{Replicas} are created, 
one for each distribution on the annealing path, and 
the chain entries in the $N$ replicas are initialized at $1,2,\ldots,N$, respectively. 
For a given \texttt{Replica}, if the \texttt{chain} number is $i$ and the \texttt{state}
is $x$, then this means that the sample corresponding to the $i$-th distribution in 
the sequence $\pi_1, \ldots, \pi_N$ is currently at location $x$.  

\medskip 
In the local exploration phase,
each \texttt{Replica}'s state is modified using an MCMC move targeting $\pi_i$,
where $i$ is given by \texttt{Replica.chain}.
The MCMC move can either modify \texttt{Replica.state} in-place, or modify the 
\texttt{Replica}'s \texttt{state} field. 
This operation is indicated by the \texttt{local\_exploration} function in 
\cref{alg:distributed_PT}.

\medskip 
In the communication phase, PT proposes swaps between pairs of replicas. 
In principle, there are two equivalent ways to do a swap.
In the first implementation, the \texttt{Replica}s 
could exchange their \texttt{state} fields.
Alternatively, they could exchange their \texttt{chain} fields.
Because we provide distributed implementations, we use the latter as it ensures that 
the amount of data that needs to be exchanged between two machines during a swap 
can be made very small (two floating point numbers), resuling in 
an exchange of $O(N)$ messages of size $O(1)$. 
Note that this cost does not vary with the dimensionality of the state space, 
in contrast to the first implementation that would transmit 
$O(N)$ messages of size $O(d)$, where $d$ is the dimension of the state space,
incurring a potentially very high communication cost for large values of $d$.

\subsection{Distributed implementation}
A distributed implementation of PT is presented in \cref{alg:distributed_PT}
and we describe the details of the implementation below.
We present the algorithm from the perspective that the number of machines available 
is equal to the number of \texttt{Replica}s and distributions, $N$. 
However, Pigeons.jl also allows for the more general case where the number of 
machines is not necessarily equal to $N$.

\subsubsection{The \texttt{PermutedDistributedArray}}
\label{sec:permuted_dist_array}
Recall that our theoretical $O(1)$ message size for communication between two machines 
is achieved by exchanging \texttt{chain} indices between \texttt{Replica}s 
instead of their \texttt{state}s.
One difficulty can be encountered in a distributed implementation, which we 
illustrate with the following simple example. 
Suppose we have $N=4$ distributions, chains, replicas, and machines, and that 
machine 1 is exploring chain 2 while machine 4 is exploring chain 3.
At one point, the \texttt{Replica} at \texttt{chain} 2 (machine 1) might need to exchange
\texttt{chain} indices with the \texttt{Replica} that has \texttt{chain} index 3 (machine 4). 
However, a priori it is not clear how machine 1 should know that it should communicate 
with machine 4 because it has no knowledge about the chain indices on the other machines.

\begin{figure*}[t] 
  \centering 
  \begin{subfigure}{0.48\textwidth}
    \centering
    \includegraphics[width=0.39\textwidth]{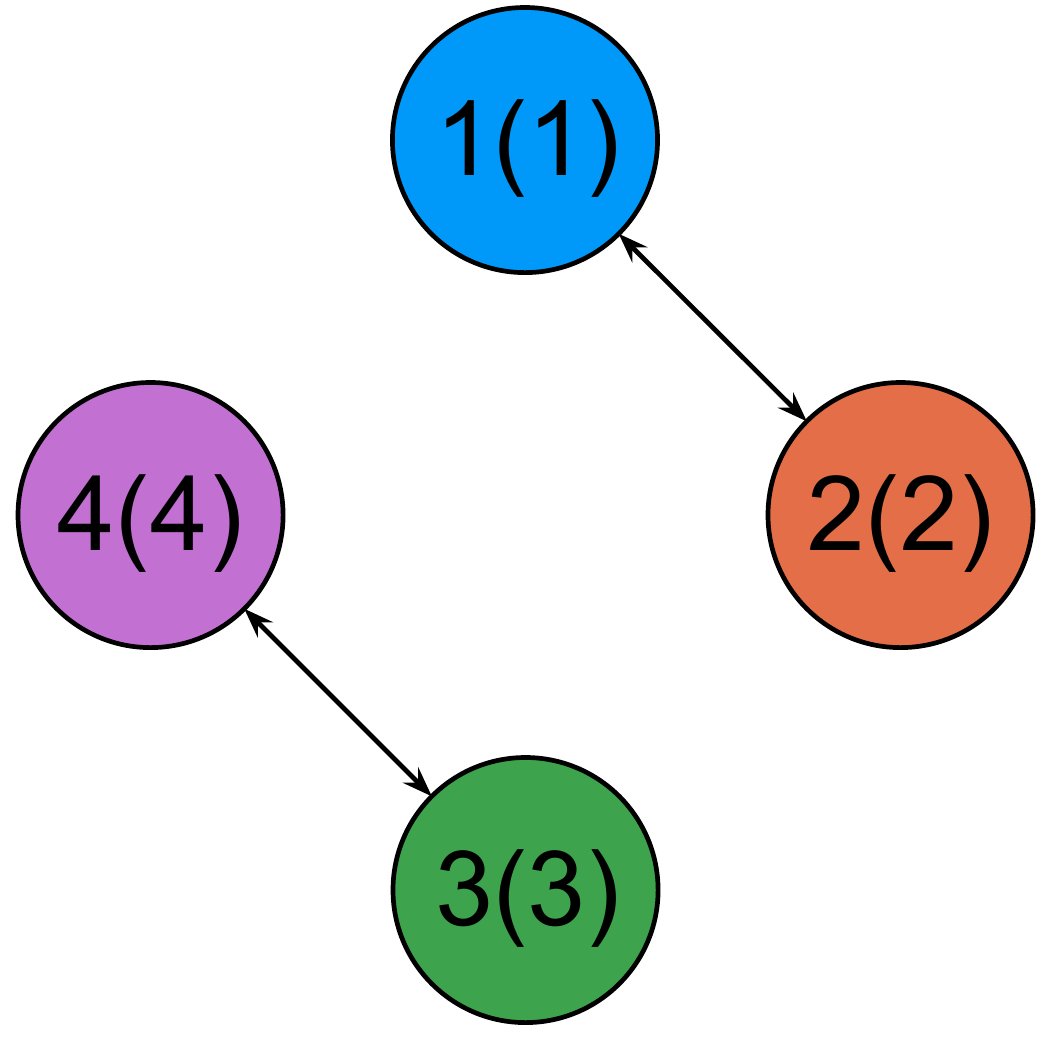}
    \includegraphics[width=0.59\textwidth]{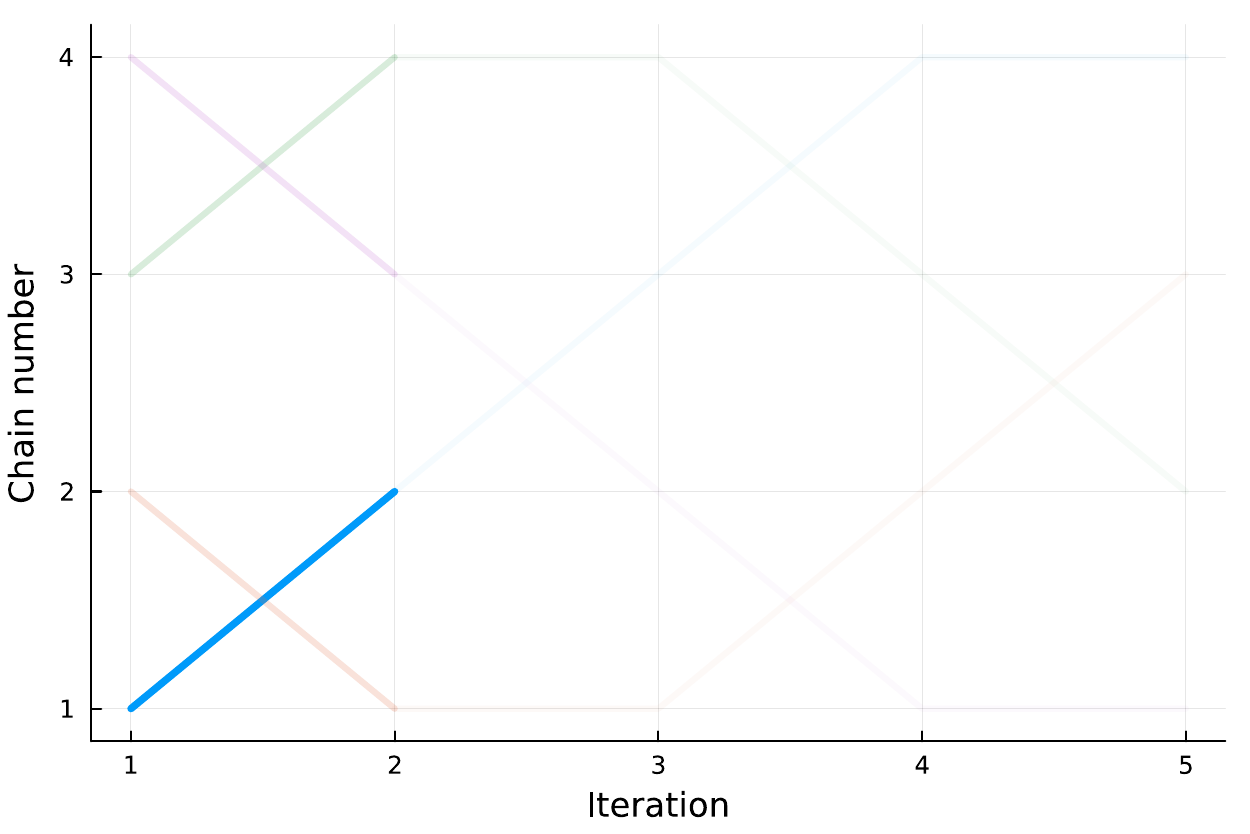}
  \end{subfigure}
  \begin{subfigure}{0.48\textwidth}
    \centering
    \includegraphics[width=0.39\textwidth]{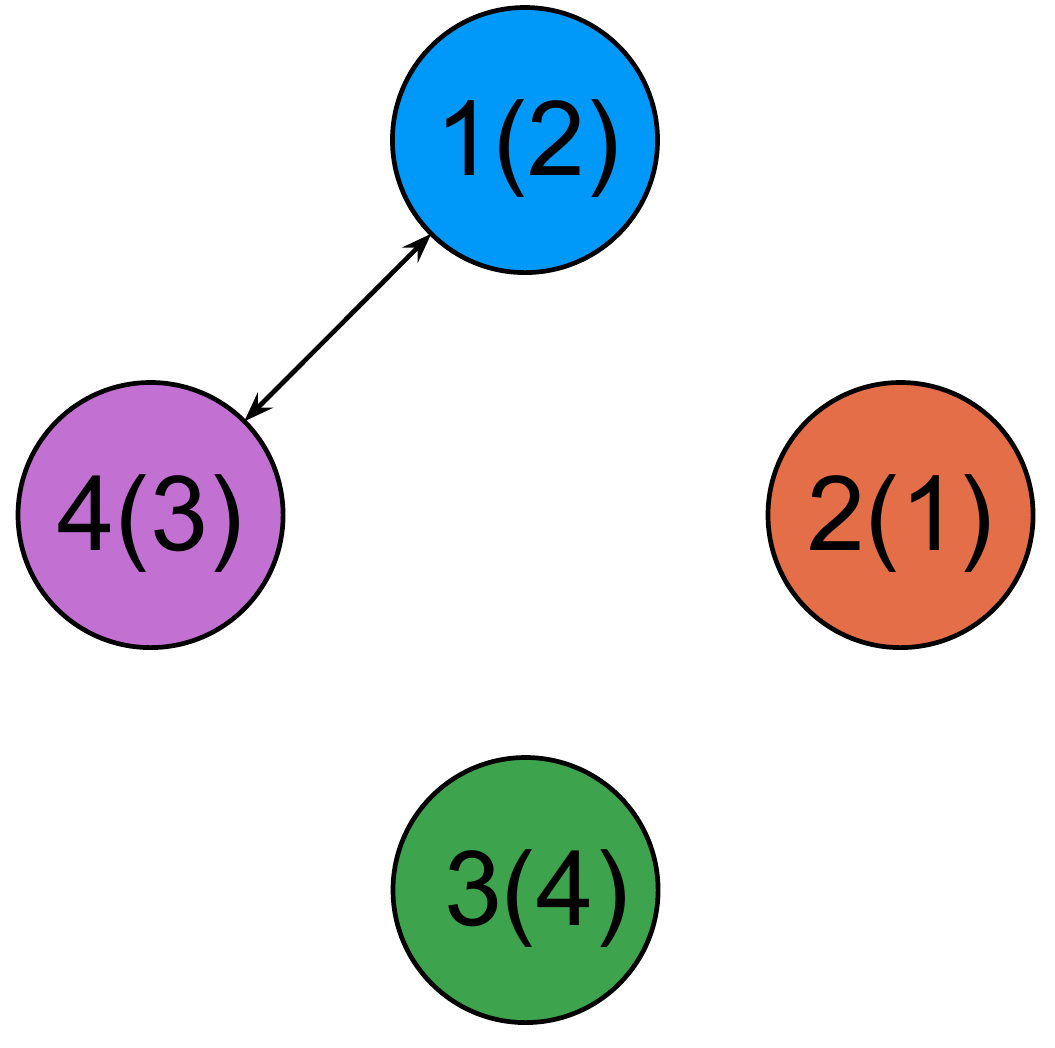}
    \includegraphics[width=0.59\textwidth]{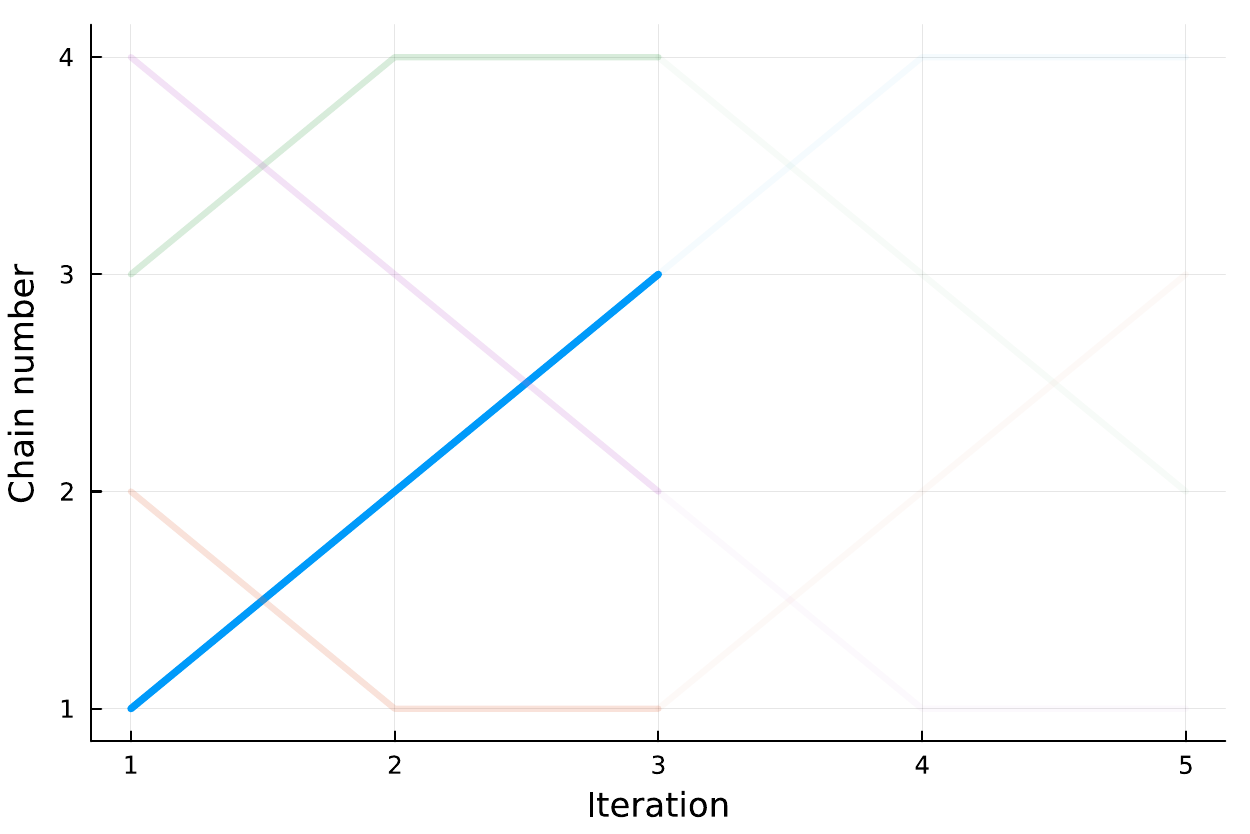}
  \end{subfigure}
  \begin{subfigure}{0.48\textwidth}
    \centering
    \includegraphics[width=0.39\textwidth]{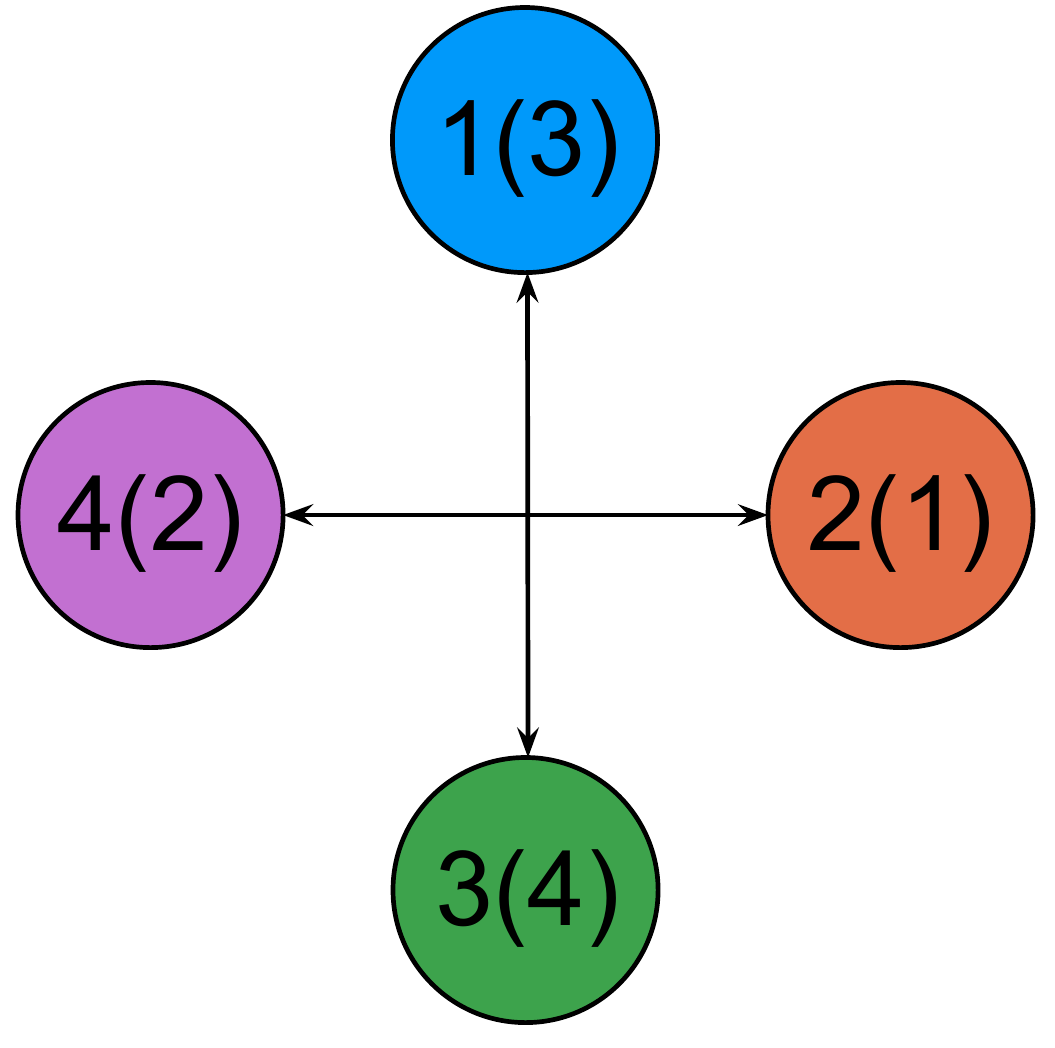}
    \includegraphics[width=0.59\textwidth]{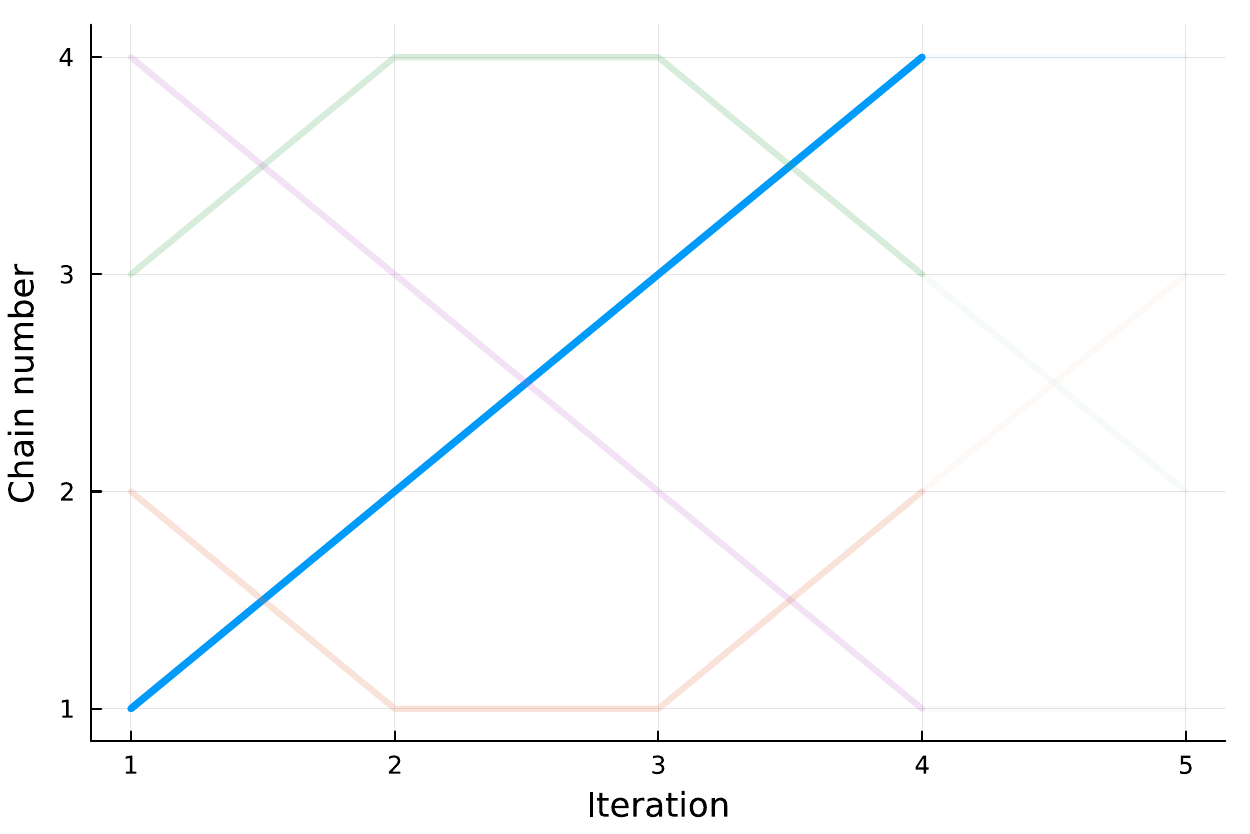}
  \end{subfigure}
  \begin{subfigure}{0.48\textwidth}
    \centering
    \includegraphics[width=0.39\textwidth]{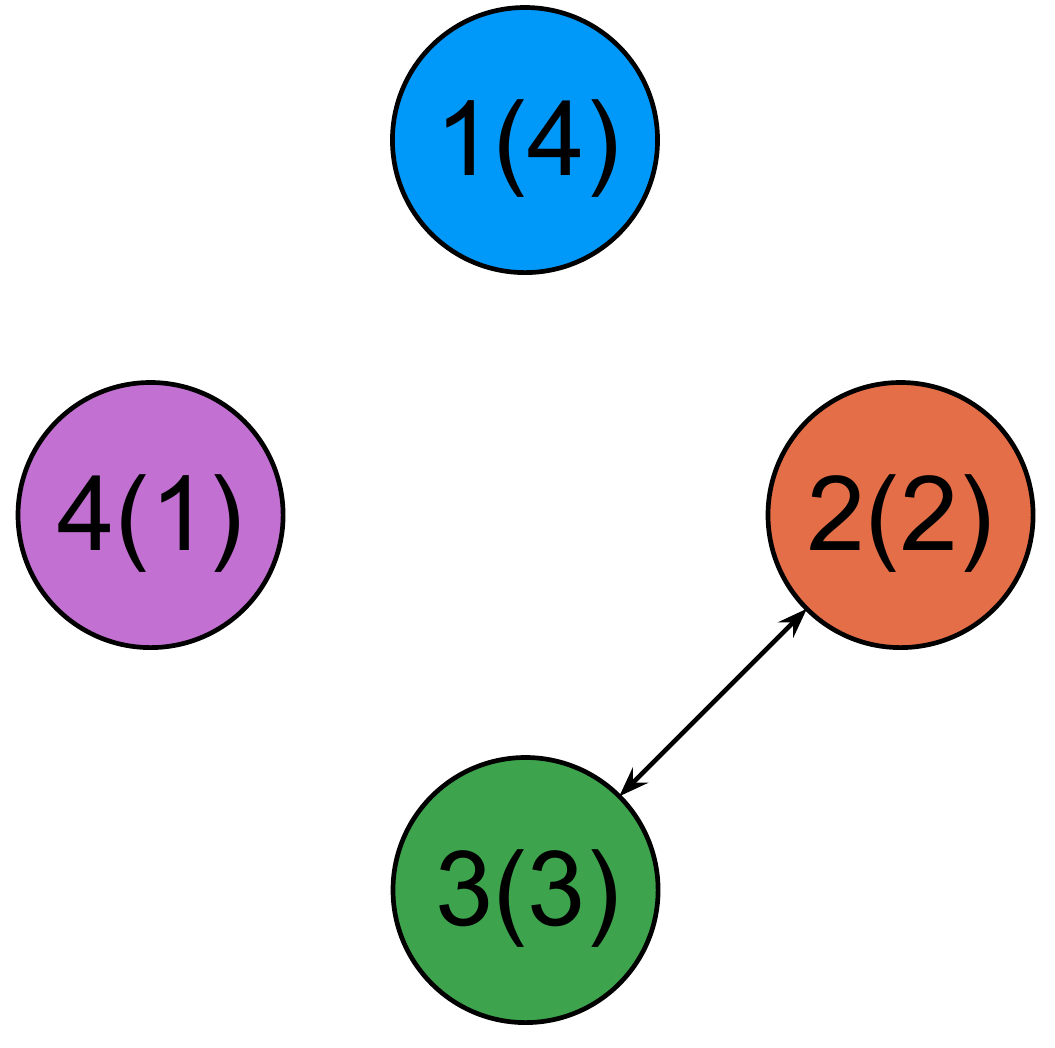}
    \includegraphics[width=0.59\textwidth]{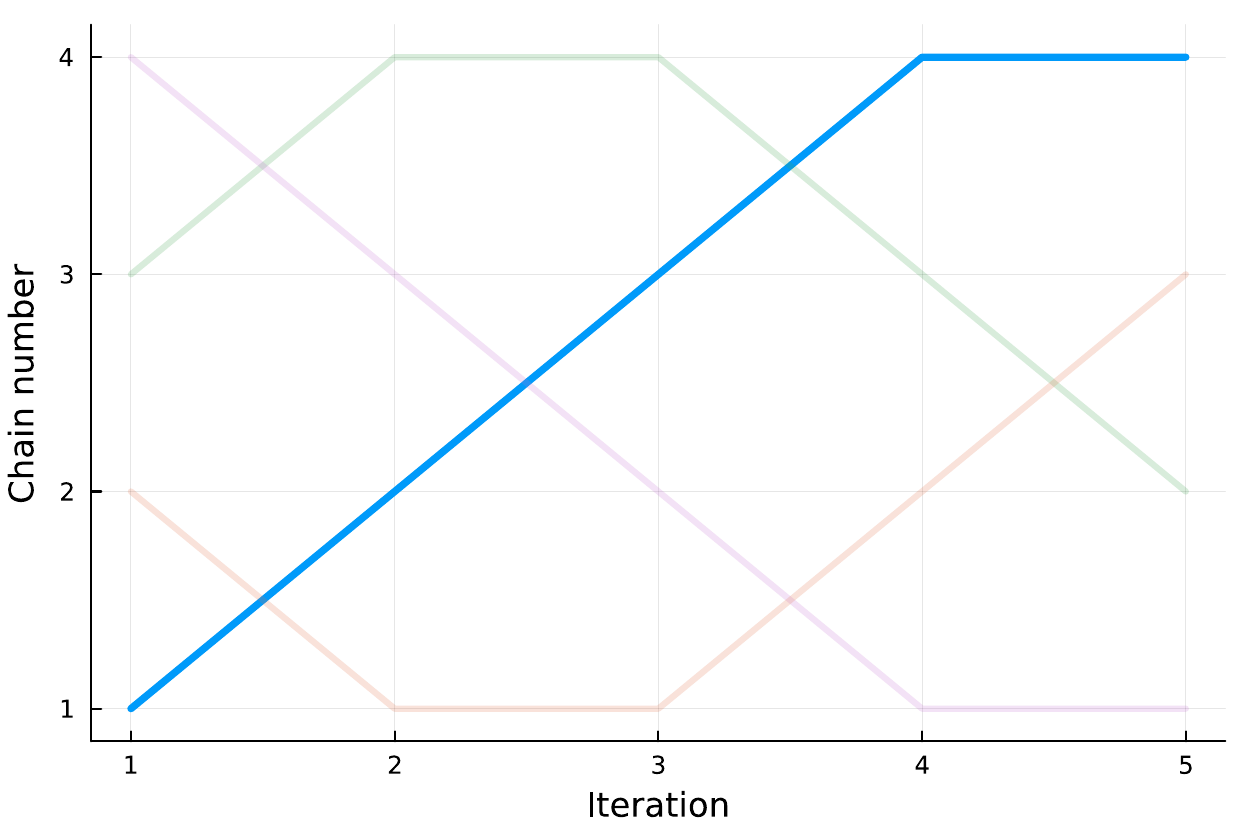}
  \end{subfigure}
  \caption{
    A summary of communication between four machines with $N=4$ PT chains.
    Circles represent machines with the machine number in the center of the circle. 
    Different colours are also used for different machines.
    Numbers in parantheses represent the \texttt{chain} 
    that is currently being explored by that machine. 
    Arrows indicate communication occuring between machines to exchange \texttt{chain} 
    indices.
    To keep track of which chains are stored on which machine, we introduce the 
    \texttt{dist\_array} of type \texttt{PermutedDistributedArray}, described in 
    \cref{sec:permuted_dist_array}, which is of length $N$. 
    The $j$-th element of \texttt{dist\_array} at a given time step indicates which 
    machine number is storing \texttt{chain} index $j$.
    The bolded curve in the figures to the right indicate the trajectory of the first 
    \texttt{Replica} over the course of each of the communication steps.      
    \textbf{Top left:} the PT replicas are initialized and the first communication step 
    is proposed. At this time step, $\texttt{dist\_array} = [1,2,3,4]$.
    \textbf{Top right:} after the first successful communication step, we now have 
    $\texttt{dist\_array} = [2,1,4,3]$.
    \textbf{Bottom left:} The second communication step is completed and 
    $\texttt{dist\_array} = [2,4,1,3]$.
    \textbf{Bottom right:} The third communication step is completed and 
    $\texttt{dist\_array} = [4,2,3,1]$.}
  \label{fig:index_process}
\end{figure*}

\medskip 
To resolve this issue, we introduce a special data structure called a 
\texttt{PermutedDistributedArray}. In the case where the number of replicas is 
equal to the number of available machines, the construction is quite simple
and effectively results in each machine storing one additional integer. 
Considering the same example above, the solution to the problem is to have
machine 1 (\texttt{chain} 2 wanting to communicate with \texttt{chain} 3) 
communicate with \textit{machine} 3. 
Machine 3 stores in its \texttt{dist\_array} variable of type 
\texttt{PermutedDistributedArray} the value 4, which is the machine number that 
stores \texttt{chain} 3. By updating these \texttt{PermutedDistributedArray} 
variables at each communication step, we can ensure that each machine $j$ is aware 
of which machine number currently stores \texttt{chain} $j$. 
Therefore, the machines act as keys in a dictionary for the communication 
permutations.
An illustration of communication between four machines is provided in 
\cref{fig:index_process}.
We note that with a \texttt{PermutedDistributedArray} we make special assumptions 
on how we access and write to the array elements. 
Several MPI processes cooperate, with each machine storing 
data for a slice of this distributed array, and at each time 
step an index of the array is manipulated by exactly one machine.

\subsubsection{MPI implementation details}
We use the MPI.jl \cite{byrne2021mpi} package to support communication between machines. 
In our pseudocode in \cref{alg:distributed_PT} we define the MPI-style 
functions \texttt{send()}, \texttt{receive!()}, 
and \texttt{waitall()}. 
The function \texttt{send()} has three arguments in our pseudocode: the first argument 
is the object to be sent, the second is the machine number to which the object should 
be sent, and finally the third argument is a unique tag for this specific send/receive 
request at this time step and on this machine, generated by \texttt{tag()}.
The \texttt{tag()} function can be any function that allows one to uniquely identify 
a tag for a given send/receive request on a given machine at any given time step $t$.
The function \texttt{receive!()} has the same last two arguments as \texttt{send()} 
except that the first argument 
specifies the object to which data should be written directly. Finally, 
\texttt{waitall(requests)} waits for all initialized MPI \texttt{requests} for a given pair 
of machines to complete.

\medskip 
In \cref{alg:distributed_PT} we also introduce the functions 
\texttt{permuted\_get()} and \texttt{permuted\_set!()}. 
Given the current machine's \texttt{dist\_array} and the 
\texttt{partner} chain with which it should communicate, 
\texttt{permuted\_get()} returns the machine number that holds the \texttt{partner} chain. 
Given the current machine's \texttt{dist\_array}, \texttt{chain} number, and machine number $j$,
\texttt{permuted\_set!()} updates the \texttt{PermutedDistributedArray} variables 
with the updated permutation.

\begin{algorithm}[t]
	\begin{algorithmic}[1]
    \Require Initial state $x_0$, sequence of distributions $\{\pi_i\}_{i=1}^N$, 
      number of iterations $T$, machine number $j$
    \State $\texttt{chain} \gets j$ \Comment{current \texttt{chain} number}
    \State $\texttt{dist\_array}[j] \gets j$ \Comment{\texttt{chain} $j$ is on machine $j$}
		
    \For{$t$ {\bf in} 1, 2, \dots, $T$}
		  \If{$t$ is even} \Comment{choose between even or odd swap}
		    \State $P \gets \{i: 1 \le i < N, i \text{ is even} \}$
		  \Else
		    \State $P \gets \{i: 1 \le i < N, i \text{ is odd} \}$
		  \EndIf

      \State $x_t \gets \texttt{local\_exploration}(\pi_\texttt{chain}, x_{t-1})$
      \State $\texttt{partner} \gets \texttt{nothing}$
      \If{$\texttt{chain} \in P$}
        \State $\texttt{partner} \gets \texttt{chain}+1$
      \ElsIf{$\texttt{chain}-1 \in P$}
        \State $\texttt{partner} \gets \texttt{chain}-1$
      \EndIf
      \If{$\texttt{partner != nothing}$ \textbf{and} \text{swap is accepted}}
        \State $\texttt{to\_machine} \gets \texttt{permuted\_get}(\texttt{dist\_array}, \texttt{partner})$
          \Comment{retrieve \texttt{dist\_array}[\texttt{partner}]}
        \State $\texttt{send\_tag} \gets \texttt{tag}(t, \texttt{chain}, j)$
        \State $\texttt{send}(\texttt{chain}, \texttt{to\_machine}, \texttt{send\_tag})$
        \State $\texttt{receive\_tag} \gets \texttt{tag}(t, \texttt{chain}, \texttt{to\_machine})$
        \State $\texttt{receive!}(\texttt{chain}, \texttt{to\_machine}, \texttt{receive\_tag})$
        \State $\texttt{waitall(requests)}$ 
          \Comment{wait for MPI requests for this pair of machines to complete}
        \State $\texttt{permuted\_set!}(\texttt{dist\_array}, \texttt{chain}, j)$
          \Comment{set value of \texttt{dist\_array}[\texttt{chain}] to \texttt{to\_machine}}
      \EndIf
      \State $c_t \gets \texttt{chain}$ 
		\EndFor
    \State \Return $\{(x_t, c_t)\}_{t=1}^T$
	\end{algorithmic}
  \caption{Distributed PT on machine $j$ (one replica per machine)}
  \label{alg:distributed_PT}
\end{algorithm}

\section{Strong parallelism invariance}
\label{sec:PI_causes}
In this section we describe potential violations of 
strong parallelism invariance (SPI) that can occur in a distributed setting. 
We also explain how we avoid these issues by using special 
distributed reduction schemes and splittable random number generators. 
Insights provided in this section can be applied 
to general distributed software beyond Julia.

\medskip 
We have identified two factors that can cause violations of our previously-defined 
SPI that standard Julia libraries 
do not automatically take care of:
\begin{enumerate}   
  \item \textbf{Non-associativity of floating point operations:} When 
  several workers perform distributed reduction of floating point values, 
  the output of this reduction will be slightly different depending on the order taken 
  during reduction. 
  When these reductions are then fed into further random operations, 
  this implies that two randomized algorithms with the same seed but using a 
  different number of workers will eventually arbitrarily diverge.

  \item \textbf{Global, thread-local, and task-local random number generators:} 
  These are the dominant approaches to parallel random number generators in current languages, 
  and an appropriate understanding of these RNGs is necessary. In particular, 
  in Julia it is important to understand the behaviour of the \texttt{@threads} macro.
\end{enumerate}
Our focus in the remainder of this section is to describe how our implementation 
solves the two above issues.

\subsection{Distributed reduction and floating point non-associativity}
After each round of PT, the machines need to exchange information such as the 
average swap acceptance probabilities, statistics to adapt a variational 
reference and adjust annealing parameters, and so on. 
For instance, suppose our \texttt{state} is univariate and real-valued and 
that each \texttt{Replica} keeps track of the number of times the target chain $N$ 
is visited as well as the mean of the univariate \texttt{state}s from the target chain.
To obtain the final estimate of the mean of the target distribution, we would like 
to pool the mean estimates from each of the \texttt{Replica}s, weighted by 
the number of times that each \texttt{Replica} visited the target chain.   
This process involves summing floating-point values that are located on each 
of the \texttt{Replica}s/machines and is an example of a reduction scheme.

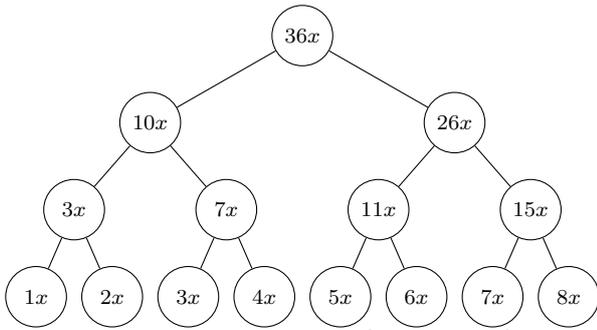
\begin{figure}[t]
  \centering
  \begin{forest}
    for tree = {circle, draw, minimum size = 0.8cm}
      [{$36x$}
        [{$10x$}
          [{$3x$}
            [{$1x$}]
            [{$2x$}]
          ]
          [{$7x$}
            [{$3x$}] 
            [{$4x$}] 
          ]
        ]
        [{$26x$}
          [{$11x$}
            [{$5x$}] 
            [{$6x$}] 
          ]
          [{$15x$}
            [{$7x$}] 
            [{$8x$}] 
          ]
        ]
      ]
  \end{forest}
  \caption{Adding eight floating point numbers $\cbra{1x, 2x, \ldots, 8x}$ 
  across eight machines. 
  Additions in each row of the tree can be performed in parallel. 
  The final result is stored in the root node of the tree and can be represented 
  by the expression $((1x + 2x) + (3x + 4x)) + ((5x + 6x) + (7x + 8x))$, indicating 
  the order of operations.}
  \label{fig:reduction_tree_8}
\end{figure}

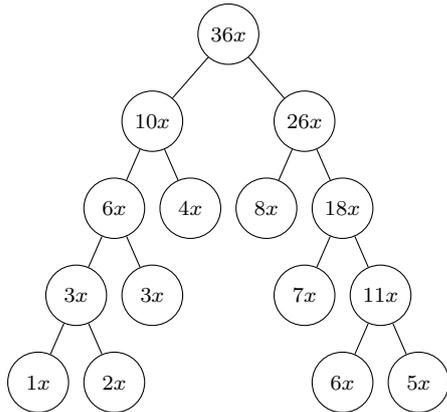
\begin{figure}[t]
  \centering
  \begin{forest}
    for tree = {circle, draw, minimum size = 0.8cm}
      [{$36x$}
        [{$10x$}
          [{$6x$}
            [{$3x$}
              [{$1x$}]
              [{$2x$}]
            ]
            [{$3x$}]
          ]
          [{$4x$}]
        ]
        [{$26x$}
          [{$8x$}]
          [{$18x$}
            [{$7x$}]
            [{$11x$}
              [{$6x$}]
              [{$5x$}]
            ]
          ]
        ]
      ]
  \end{forest}
  \caption{One possible way of adding eight floating point numbers 
  $\cbra{1x, 2x, \ldots, 8x}$ across two machines. 
  The final result is stored in the root node of the tree and can be represented 
  by the expression $(((1x + 2x) + 3x) + 4x) + (((5x + 6x) + 7x) + 8x)$.
  Note that the order of operations in this expression is different from the one 
  presented in \cref{fig:reduction_tree_8}.}
  \label{fig:reduction_tree_2}
\end{figure}

\medskip 
To illustrate why distributed reduction with floating point values can violate 
strong parallelism invariance if not properly implemented, we consider the 
following toy example. 
Suppose we have 8 machines storing the floating point numbers $\cbra{1x, 2x, \ldots, 8x}$, 
as illustrated in \cref{fig:reduction_tree_8}, where we use $x = 10 e^1 \approx 27.1828$
in the following examples.
In this case, if our reduction procedure is to sum the floating point numbers, 
we know that our final answer should be approximately $36x$. 
However, depending on the exact order in which floating point addition is carried out, 
the answers might not all be the same and exactly equal to $36x$.
For instance, in \cref{fig:reduction_tree_8} we see that the order of operations 
for eight machines would be given by 
\[
  ((1x + 2x) + (3x + 4x)) + ((5x + 6x) + (7x + 8x)) \\
  \approx 978.5814582452562.
\]
In contrast, with two machines, one possible order of operations might be 
\[
  (((1x + 2x) + 3x) + 4x) + (((5x + 6x) + 7x) + 8x) \\
  \approx 978.5814582452563.
\]
A possible reduction tree for two machines is illustrated in \cref{fig:reduction_tree_2}.

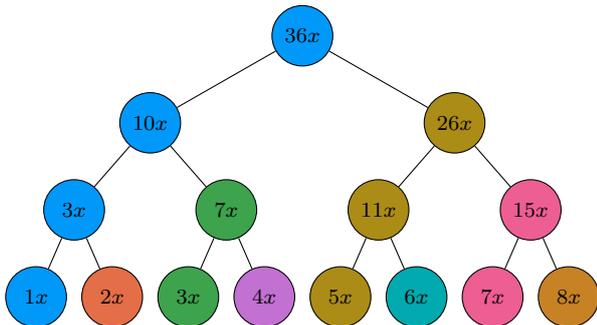
\begin{figure}[t]
  \centering
  \begin{forest}
    for tree = {circle, draw, minimum size = 0.8cm}
      [{$36x$}, fill = julia1
        [{$10x$}, fill = julia1
          [{$3x$}, fill = julia1
            [{$1x$}, fill = julia1]
            [{$2x$}, fill = julia2]
          ]
          [{$7x$}, fill = julia3
            [{$3x$}, fill = julia3] 
            [{$4x$}, fill = julia4] 
          ]
        ]
        [{$26x$}, fill = julia5
          [{$11x$}, fill = julia5
            [{$5x$}, fill = julia5] 
            [{$6x$}, fill = julia6] 
          ]
          [{$15x$}, fill = julia7
            [{$7x$}, fill = julia7] 
            [{$8x$}, fill = julia8] 
          ]
        ]
      ]
  \end{forest}
  \caption{Addition of eight floating point numbers across eight machines with a 
  guarantee on SPI. Each machine is represented by a different colour. 
  The final result can be represented 
  by the expression $((1x + 2x) + (3x + 4x)) + ((5x + 6x) + (7x + 8x))$.}
  \label{fig:reduction_tree_colour_8}
\end{figure}

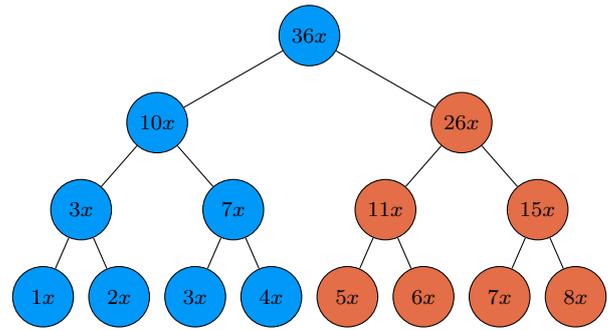
\begin{figure}[t]
  \centering
  \begin{forest}
    for tree = {circle, draw, minimum size = 0.8cm}
      [{$36x$}, fill = julia1
        [{$10x$}, fill = julia1
          [{$3x$}, fill = julia1
            [{$1x$}, fill = julia1]
            [{$2x$}, fill = julia1]
          ]
          [{$7x$}, fill = julia1
            [{$3x$}, fill = julia1] 
            [{$4x$}, fill = julia1] 
          ]
        ]
        [{$26x$}, fill = julia2
          [{$11x$}, fill = julia2
            [{$5x$}, fill = julia2] 
            [{$6x$}, fill = julia2] 
          ]
          [{$15x$}, fill = julia2
            [{$7x$}, fill = julia2] 
            [{$8x$}, fill = julia2] 
          ]
        ]
      ]
  \end{forest}
  \caption{Addition of eight floating point numbers across two machines with a 
  guarantee on SPI. Each machine is represented by a different colour. 
  The final result can be represented 
  by the expression $((1x + 2x) + (3x + 4x)) + ((5x + 6x) + (7x + 8x))$,
  which is the same as that given by eight machines in \cref{fig:reduction_tree_colour_8}.}
  \label{fig:reduction_tree_colour_2}
\end{figure}

\medskip 
To avoid the issue of non-associativity of 
floating point arithmetic, we ensure that the order in which operations are performed 
is exactly the same, irrespective of the number of processes/machines and threads. 
This is achieved by making sure that every value to be added---if addition is our 
reduction operation---is a leaf node in the reduction tree, irrespective of the 
number of machines available to perform the reduction. 
For instance, if $N$ values are to be reduced, then the reduction tree would have 
$N$ leaf nodes. If $M$ machines are available, these machines are then assigned 
in such a way that the order of operations is as if there were $N$ machines available. 
To do so, it may be necessary for a machine to ``communicate with itself'', 
imitating the behaviour that would be present if there were $N$ machines available.
\cref{fig:reduction_tree_colour_8} and \cref{fig:reduction_tree_colour_2} 
illustrate the reduction procedure for eight and two machines, respectively.

\subsection{Splittable random streams}
\label{sec:splittable_randoms}
Another building block towards achieving SPI is a \emph{splittable random stream}
\cite{lecuyer1988splittable,burton1992splittable}. 
Julia uses \emph{task-local} random number generators, a notion that is 
related to (but does not necessarily imply) strong parallelism invariance. 
A \emph{task} is a unit of work on a machine.
A task-local RNG would then mean that a separate RNG is 
used for each unit of work, hopefully implying strong parallelism invariance
if the number of tasks is assumed to be constant. 
Unfortunately, this is not the case when a separate task is created for each thread 
of execution in Julia.
We note that the \texttt{@threads} macro in Julia creates \texttt{nthreads()}
tasks and thus \texttt{nthreads()} pseudorandom number generators.
This can break strong parallelism invariance as the output may depend on the number of 
threads.

\medskip 
To motivate splittable random streams, consider the following toy example 
that violates our notion of SPI:
\begin{lstlisting}[language = Julia]
using Random
import Base.Threads.@threads

println("Num. of threads: $(Threads.nthreads())")

const n_iters = 10000
result = zeros(n_iters)
Random.seed!(1)
@threads for i in 1:n_iters
    result[i] = rand()
end
println("Result: $(last(result))")
\end{lstlisting}
With eight threads, this outputs:
\begin{lstlisting}[language = Julia]
Num. of threads: 8
Result: 0.25679999169092793
\end{lstlisting}
Julia guarantees that if we rerun this code, as long as we are using eight threads, 
we will always get the same result, irrespective of the multi-threading 
scheduling decisions implied by the \texttt{@threads}-loop. 
Internally, Julia works with task-local RNGs and the \texttt{@threads} macro 
spawns \texttt{nthreads()} number of task-local RNGs.
For this reason, with a different number of threads, the result is different:
\begin{lstlisting}[language = Julia]
Num. of threads: 1
Result: 0.8785201210435906
\end{lstlisting}

In this simple example above, the difference in output is perhaps not too concerning, 
but for our parallel tempering 
use case, the distributed version of the algorithm is significantly more complex 
and difficult to debug compared to the single-threaded one. We therefore take task-local 
random number generation one step further and achieve SPI, which 
guarantees that the output is not only reproducible with respect to repetitions 
for a fixed number of threads, but also for different numbers of threads or processes.

\medskip 
A first step to achieve this is to associate one random number 
generator to each \texttt{Replica}. 
To do so, we use our SplittableRandoms.jl package, which is a Julia implementation 
of Java SplittableRandoms. Our package offers an implementation of  
SplitMix64 \cite{steele2014fast},
which allows us to turn one seed into an arbitrary collection of pseudo-independent 
RNGs. 
A quick example of how to use the SplittableRandoms.jl library is given below. 
By splitting a master RNG using the \texttt{split()} function, 
we can achieve SPI even with the use of the \texttt{@threads} macro. 

\begin{lstlisting}[language = Julia]
using Random
using SplittableRandoms: SplittableRandom, split
import Base.Threads.@threads

println("Num. of threads: $(Threads.nthreads())")

const n_iters = 10000
const master_rng = SplittableRandom(1)
result = zeros(n_iters)
rngs = [split(master_rng) for _ in 1:n_iters]
Random.seed!(1)
@threads for i in 1:n_iters
    result[i] = rand(rngs[i])
end
println("Result: $(last(result))")
\end{lstlisting}
With one and eight threads, the code above outputs
\begin{lstlisting}[language = Julia]
Num. of threads: 1
Result: 0.4394633333251359
\end{lstlisting}
\begin{lstlisting}[language = Julia]
Num. of threads: 8
Result: 0.4394633333251359
\end{lstlisting}
\section{Related work}
Automated software packages for Bayesian inference have revolutionized Bayesian data 
analysis in the past two decades, and are now a core part of a typical applied statistics 
workflow. For example, packages such as BUGS 
\cite{lunn2013bugs, lunn2009bugs, lunn2000winbugs}, 
JAGS \cite{hornik2003jags}, Stan \cite{carpenter2017stan}, PyMC3 
\cite{salvatier2016probabilistic}, and Turing.jl \cite{ge2018turing} have been widely used in 
many scientific applications.

\medskip 
These software packages often provide two key user-facing components: (1) a 
probabilistic programming language (PPL), which allows users to specify Bayesian 
statistical models in code with a familiar, mathematics-like syntax; and (2) an 
inference engine, which is responsible for performing computational Bayesian
inference once the model and data have been specified. 
Pigeons.jl focuses on the development 
of a new inference engine that employs distributed, high-performance computing.

\medskip
Inference engines available in existing software packages are varied in their 
capabilities and limitations. For instance, the widely-used Stan inference engine 
is only capable of handling real-valued (i.e., continuous) parameters. 
Other tools, such as JAGS, are capable of handling discrete-valued parameters but 
are limited in their capability to handle custom data-types 
(e.g. phylogenetic trees). 
Of those that have the capability to model arbitrary data types, 
none have an automatically distributed implementation to our knowledge.
Turing.jl \cite{ge2018turing} offers another popular inference engine and PPL, 
however Pigeons.jl allows one to interface with several different PPLs as well as 
to easily perform distributed computation.

\medskip
There is also a vast literature on distributed and parallel 
Bayesian inference algorithms 
\cite{bardenet2017markov, brockwell2006parallel, calderhead2014general,
jacob2020unbiased, jacob2011using, lee2010utility, scott2016bayes, 
wang2015parallelizing, wu2017average, zhu2017big}.
These methods unfortunately do not lead to widely 
usable software packages because they either introduce unknown amounts of 
approximation error, involve significant communication cost, or reduce the 
generality of Bayesian inference. 

\section{Conclusion}
Pigeons.jl is a Julia package that enables users with no experience in distributed 
computing to efficiently approximate posterior distributions and solve challenging 
Lebesgue integration problems over a distributed computing environment.
The core algorithm behind Pigeons.jl is distributed, non-reversible parallel tempering 
\cite{syed2021nrpt,surjanovic2022vpt}. Pigeons.jl can be used in a multi-threaded context, 
as well as distributed over up to thousands of MPI-communicating machines. 
Further, Pigeons.jl is designed so that for a given seed, the output is \textit{identical}
regardless of the number of threads or processes used.

\section{Acknowledgements}
NS acknowledges the support of a Vanier Canada Graduate Scholarship. 
PT acknowledges the support of the Black Hole Initiative at Harvard University, 
which is funded by grants from the John Templeton Foundation and the 
Gordon and Betty Moore Foundation to Harvard University.
PT also acknowledges support by the National Science Foundation grants AST-1935980 
and AST-2034306 and the Gordon and Betty Moore Foundation (GBMF-10423).
SS acknowledges the support of EPSRC grant EP/R034710/1 CoSines.
ABC and TC acknowledge the support of an NSERC Discovery Grant.
We also acknowledge use of the ARC Sockeye computing platform from the
University of British Columbia, as well as cloud computing resources provided 
by Oracle.


\bibliographystyle{juliacon}
\bibliography{ref.bib}

\end{document}